\documentclass{aa}

\usepackage{graphicx}
\usepackage{txfonts}
\usepackage{subfigure}
\usepackage{natbib}
\usepackage{color}
\begin{document}
\newcommand{\blue}{\textcolor{blue}}
\newcommand{\red}{\textcolor{red}}

\title{Scattered moonlight observations with X-Shooter}
\subtitle{Implications for the aerosol properties at Cerro Paranal and the ESO sky background model\thanks{Based on observations collected at the European Southern Observatory under ESO programme 491.L-0659.}}

\author{A. Jones
          \inst{1,2}
\and
          S. Noll\inst{3,4,1}
\and
	  W. Kausch\inst{1}
\and
	  S. Unterguggenberger\inst{1}
\and
	  C. Szyszka\inst{1}
\and
	S. Kimeswenger\inst{1,5}
          }

   \institute{Institut f\"ur Astro- und Teilchenphysik, Universit\"at Innsbruck, Technikerstr. 25, Innsbruck A-6020, Austria
   \and
	Physics and Astronomy Department, University of Alabama, 514 University Blvd, Tuscaloosa, AL 35487 USA\\
	\email{amjones37@ua.edu}
   \and
   Institut f\"ur Physik, Universit\"at Augsburg, Universit\"atsstra\ss e 1, Augsburg 86159, Germany
   \and
   Deutsches Fernerkundungsdatenzentrum, Deutsches Zentrum f\"ur Luft- und Raumfahrt, M\"unchener Stra\ss e 20, We\ss{}ling-Oberpfaffenhofen 82234, Germany
  \and
   Instituto de Astronom{\'{i}}a, Universidad Cat{\'{o}}lica del Norte,
                Avenida Angamos 0610, Antofagasta, Chile
          }

   \date{}

\abstract{Estimating the sky background is critical for ground-based astronomical research.  In the optical, scattered moonlight dominates the sky background, when the moon is above the horizon.  The most uncertain component of a scattered moonlight model is the aerosol scattering.  The current, official sky background model for Cerro Paranal uses an extrapolated aerosol extinction curve.  With a set of X-Shooter sky observations, we have tested the current sky background model as well as determined the aerosol extinction from the ultra-violet to near-infrared (NIR).  To our knowledge, this is the first time that a scattered moonlight model has been used for this purpose.  These observations were taken of blank sky, during three different lunar phases, and at six different angular distances from the moon for each night/lunar phase.   Overall, the current model does reproduce the observations for average conditions decently well.  Using a set of sky background models with varying aerosol distributions to compare with the observations, we found the most likely aerosol extinction curves, phase functions, and volume densities for the three nights of observations and compare them with the current model.  While there were some degeneracies in the aerosol scattering properties, the extinction curves tend to flatten towards redder wavelengths and are overall less steep compared to the extrapolated curve used in the current model.  Also, the current model had significantly less coarse particles compared to the favored volume densities from the X-Shooter data.  Having more coarse particles affects the phase function by being more peaked at small angular distances.  For the three nights of sky observations, the aerosol size distributions differed, most likely reflecting the changes in atmospheric conditions and aerosol content, which is expected.  In short, the current sky background model is in fair agreement with the observations, and we have determined better aerosol extinction curves and phase functions for Cerro Paranal.  Using nighttime sky observations of scattered moonlight and a set of sky background models is a new method to probe the aerosol content of the atmosphere.
}

\keywords{Atmospheric effects - Radiative transfer - Scattering - Methods: data analysis - Methods: observational - Techniques: spectroscopic}
\maketitle

\section{Introduction}

Ground-based astronomy is limited by how well the sky background can be estimated and removed \citep[see e.g.][]{1989PASP..101..306G,1998A&AS..127....1L}.  Many times the target observation and the sky are observed simultaneously or sequentially in order to obtain temporally and spatially a decent approximation of the sky at the time and location of the target observation.  However, sometimes this is not possible and a sky background model is needed.  Also, a sky background model is useful for estimating how long an observation should be in order to obtain the desired signal-to-noise ratio required for the given science case.  Therefore for modern, ground-based telescopes, an accurate sky background model and characterization of the night sky is necessary.  The sky background consists of many components that contribute to both radiance and transmission spectra.  Some of these components include direct and indirect zodiacal light, scattered moonlight, scattered starlight, and absorption, emission and scattering by particles in the atmosphere (troposphere, stratosphere, and the upper atmosphere) \citep[e.g.][and references therein]{1967ApJ...147..255W,1975A&A....39..325S,1998A&AS..127....1L,2010ApJ...720..811B,2012A&A...543A..92N}.  These components can vary temporally and spatially on different scales.

For the European Southern Observatory (ESO) a sky background model was developed for the purpose of the exposure time calculator (ETC) for the Very Large Telescope (VLT) at Cerro Paranal and the future site of the Extremely Large Telescope (ELT) at Cerro Armazones \citep{2012A&A...543A..92N,2013A&A...560A..91J}.   Both sites are located in the Atacama desert of Chile at high elevations, 2\,635 and 3\,046\,km for the VLT and ELT, respectively.  One of the reasons for choosing these locations is that the atmosphere is quite dry and clean (i.e. there are small amounts of water vapor and particles).
 
A major component of the sky background model is scattered moonlight, which dominates the night sky brightness in the optical when the moon is above the horizon.  Surprisingly, there have not been many studies of scattered moonlight.  \citet{1987W} published a table of the sky brightnesses taken at Cerro Tololo, Chile during 5 different lunar phases in the U, V, B, R, and I-bands.  This was used by many observatories to estimate the scattered moonlight for a couple of decades, but there is no information about the position of the moon or patch of sky observed.  Then \citet{1991PASP..103.1033K} designed an empirical model for the brightness of the moon, using V-band data taken at Mauna Kea, Hawaii.  This model depended on the lunar phase, positions of the moon and target, as well as atmospheric conditions, and had an uncertainty between 8 and 23\%.  \citet{2012A&A...543A..92N} extrapolated this empirical model from being only photometric to being spectroscopic using a set of observations from the UV FOcal Reducer and low dispersion Spectrograph \citep[FORS1,][]{1998Msngr..94....1A,2008A&A...481..575P} at the VLT.  It used the solar spectrum from \citet{1996AJ....112..307C} and considered the wavelength dependence of Rayleigh scattering as well as from aerosols using the aerosol extinction curve from \citet{2011A&A...527A..91P}.  Aerosols are particles that are larger than a typical molecule and can scatter light in the atmosphere, usually in the troposphere and stratosphere.  This extended model was part of the improved ESO sky background model for the ETC, where the Walker table was replaced by this model.  However, this extended model was an empirical fit with many parameters.  \citet{2013A&A...560A..91J} used a new method for estimating the scattered moonlight was developed to be used for the ESO ETC and other applications.  We followed the path of light from the sun, to being reflected off the moon, to being scattered in the Earth's atmosphere before entering the telescope.  The solar spectrum was also taken from \citet{1996AJ....112..307C} and the albedo was interpolated from \citet{2005AJ....129.2887K}.  An improved set of scattering calculations are described by \citet{2013A&A...560A..91J} and includes Rayleigh scattering from molecules as well as aerosol scattering.  \citet{2013A&A...560A..91J} used the aerosol extinction curve, which gives the optical depth from aerosols due to scattering and absorption as a function of wavelength, for Cerro Paranal from \citet{2011A&A...527A..91P}, like the previous extended ESO model.  For the aerosol scattering phase function, we decomposed this extinction curve into typical different background aerosol particles from \citet{2012acc.book.....W} using the Mie calculations from \citet{1983asls.book.....B,2004ApOpt..43.5386G}, which determined the phase function.  The aerosol phase function describes the angular distribution of light reflected off an aerosol particle as a function of wavelength.  Aerosols tend to preferentially scatter light forward.  The phase function has no physical units and its integral over all scattering directions is normalized to $4\pi$\,sr.   There were a couple of problems with this method, such as the aerosol extinction curve was determined between 0.4 and 0.8\,$\mu$m and there were several degeneracies amongst the different aerosol particles.  However, in order to characterize the amount of moonlight expected in a given observation, knowing the amount of each type of aerosols present is critical for calculating the scattered moonlight.  The aerosol scattering is the most uncertain part of the scattered moonlight model.  To better characterize the aerosol distribution and hence improve the scattered moonlight model, a set of X-Shooter observations \citep{2011A&A...536A.105V} was obtained at different lunar phases and distances from the moon.  X-Shooter is an echelle spectrograph at the VLT and can observe simultaneously from 0.3 to 2.1\,$\mu$m (with a K-blocking filter).  

With the X-Shooter observations we can validate the sky background model, with an emphasis on the scattered moonlight model, using a set of dedicated blank sky observations to probe different aspects of the model.  The scattered moonlight model was previously tested using FORS1 data only in the optical range and with these new observations it can be independently tested for the three nights of observations.  Because the three nights have a range in lunar phases and each night also has observations at different angular distances to the moon, we can verify that the model accurately depends on these parameters, lunar phase and angular distance.  Additionally with the X-Shooter data set, we can probe the aerosol size distribution for each night of observations.  To our knowledge, this is the first time a scattered moonlight model has been used for this purpose. The full wavelength range and the different angular distances, also provide insight into the coarse aerosol particles, which are difficult to study with only an optical extinction curve.  In light of the previous lack of constraints, the measurements of coarse particles on all three nights, even though the aerosol properties vary on short and long timescales, will help to better match the scattered moonlight model with the X-Shooter observations.

The paper is organized as follows.  In Section 2 we describe the X-Shooter observations used in this study.  The analysis is discussed in Section 3, including the current sky background model, how we measure the aerosol extinction and the aerosol grid used.  In Section 4 we show how well the current scattered moonlight compares with the observations.  The results for the aerosol properties are described in Section 5.  Section 6 is a discussion and Section 7 has the conclusions.  Additionally, there is Appendix A which provides values for the scattered moonlight brightness in the J-band.

\section{Data}

\begin{figure*}[!ht]
   \centering
   \includegraphics[width=0.95\textwidth]{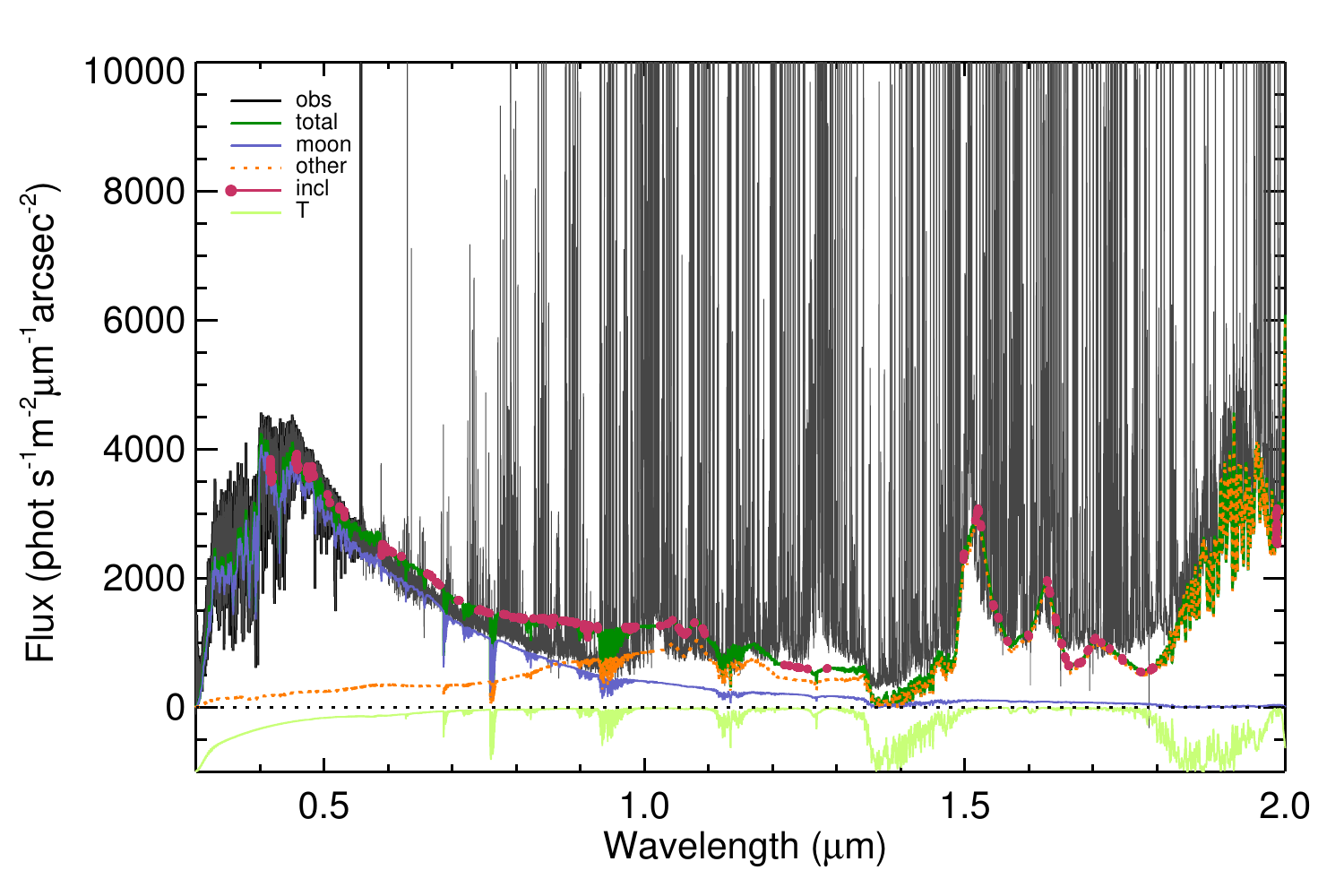}
   \caption{An example spectrum for the scattered moonlight determination.  The observed spectrum (\textit{dark gray}) was taken during Run B at $\rho=45^\circ$.  Overlaid is an example of the current total sky background model (\textit{green}) with the sky lines removed for clarity.  Also shown are the inclusion regions for the analysis (\textit{pink filled circles}).  The scattered moonlight model (\textit{blue}) and the other model components (\textit{orange dotted}), except the sky lines, are also shown.  Below the dotted black line is the transmission curve (\textit{light green}).}
\label{eg_spec}
    \end{figure*}

Our observations were taken with X-Shooter spectrograph \citep{2011A&A...536A.105V} at the VLT in Cerro Paranal, Chile with the proposal ID 491.L-0659.  The observations were of blank or plain sky, i.e. observations without any astronomical source in the field and only of the sky background, taken at six different angular distances $\rho$ (7, 13, 20, 45, 90, and 110$^\circ$) from the moon on three different nights (Run A, B, and C).  The nights were chosen to have different lunar phases, one within three days of full moon (Run A), one three to five days from full moon (Run B), and another with an additional three to five days from the full moon so close to 50\% illumination (Run C).  X-Shooter observes simultaneously with three arms (UVB, VIS, and NIR) to cover 0.3 to 2.1\,$\mu$m.  We used the K-blocking filter to minimize the amount of scattered light from the K-band, hence the observations only extend to 2.1 instead of 2.5\,$\mu$m.  For the sky observations, the slit widths were 1.6, 1.5, and 0.8\;arcsecs for the UVB, VIS, and NIR arms, respectively, and the length was 11\;arcsecs.  According to the ESO archive, the moon illumination was 100, 97, and 56\% for Run A, B, and C, respectively.   More details about the observations are shown in Table \ref{tab_obs}, including the run, target observation, date and time, right ascension (RA), declination (dec), exposure (exp) time for the three arms, airmass (column density of air relative to the zenith), and seeing conditions.  These values are all taken from the ESO archive.  Run A and C were both taken in April, while Run B was observed in July.  Run A also had some clouds towards the end of the run, which were noted in the observing logs.  An example of one of the observations during Run B at $\rho=45^\circ$ is shown in Fig. \ref{eg_spec}.

Along with the plain sky observations, spectrophotometric standard stars were also observed each night at two different airmasses.  The star was observed before and after the sky observations.  LTT 7987 and LTT 3218 were used as spectrophotometric standard stars.  For the stars, we used a slit width of 5.0\;arcsecs for all three arms.  For more details, see Table \ref{tab_obs}.  The reason for observing the standard stars at two different airmasses was to determine the aerosol extinction curve using the so-called Langley method.  The Langley method involves looking at the ratio of the fluxes for a standard star observed at two different airmasses and assuming that the change in flux is only due to changes in atmospheric extinction \citep{aerosol_2013}.

The sky observations allowed us to check the dependencies of the sky brightness on both the lunar phase and the angular distance $\rho$ from the moon as a function of wavelength. By angular distance we are referring to the apparent distance between the target observation and the moon in the night sky measured in degrees.  The observations with a higher fractional lunar illumination are good for tests of the scattered moonlight model, especially the ones at small $\rho$, whereas the observations at lower fractional lunar illumination and large $\rho$ should have very little scattered moonlight and the other components of the sky model can be studied.  Additionally, at bluer wavelengths Rayleigh scattering dominates over aerosol scattering due to a steeper wavelength dependence of $\sim\lambda^{-4}$ versus $\sim\lambda^{-1}$, respectively.  At the redder wavelengths, airglow lines and continuum are the main source of sky brightness.

The observations were reduced using the ESO X-Shooter reflex pipeline version 2.3.0.  The flux calibration was improved by using the method described by \citet{2015ACP....15.3647N}.  This method relies on instrument response curves that were derived from standard star spectra taken between October 2009 to March 2013.  The latest response curve was used for the flux calibration of our observations, and only an error of less than a few percent is expected.  The other change from the standard pipeline for the sky observations was the 1D extraction.  From the 2D sky observations we took the median value of flux for each wavelength pixel to obtain the 1D spectra.  The stars were observed in nodding mode, hence they were located at the edge of the slit.  Due to atmospheric differential refraction (ADR), different proportions of the flux were refracted out of the slit as a function of wavelength.  For the calibration of the standard stars, they were corrected for any loss of flux along the slit, by modeling how the star moved throughout each observation as a function of wavelength.  However the flux in the standard stars is still uncertain at the few percent level. 

\begin{table*}
\caption{X-Shooter observations \label{tab_obs}}
\centering
\begin{tabular}{clccclcc}
\hline\hline
\noalign{\smallskip}
run & object & date and time & RA & dec & exp time (s) & airmass & seeing (DIMM)\tablefootmark{a} \\
 & & & hh:mm:ss.ss & dd:mm:ss.s & UVB,\,VIS,\,NIR & &arcsec  \\
\noalign{\smallskip}
\hline
\noalign{\smallskip}
A&	LTT 7987    &2013-04-26 06:00:46      &20:10:56.33   & -30:13:11.5     &115,\,183,\,270                 &2.02    &0.75\\
A&	sky 7 deg   &2013-04-26 06:16:36      &14:35:00.00   & -08:58:57.0     &629,\,540,\,600                 &1.09    &0.75\\
A&	sky 13 deg  &2013-04-26 06:32:33      &14:35:03.00   & -03:00:33.0     &629,\,540,\,600                 &1.17    &0.73\\
A&	sky 20 deg  &2013-04-26 06:58:43      &15:34:55.00   & +00:00:30.0     &629,\,540,\,600                 &1.14    &0.82\\
A&	sky 45 deg  &2013-04-26 07:15:09      &16:59:58.00   & +15:00:03.0     &629,\,540,\,600                 &1.30    &0.80\\
A&	sky 90 deg\tablefootmark{b}  &2013-04-26 08:45:59      &23:30:00.00   & -64:59:20.0     &629,\,540,\,600                 &2.15    &0.91\\
A&	sky 110 deg\tablefootmark{b} &2013-04-26 09:30:03      &23:30:00.00   & -29:59:11.0     &629,\,540,\,600                 &1.89    &1.06\\
A&	LTT 7987    &2013-04-26 09:47:42      &20:10:56.33   & -30:13:09.8     &115,\,183,\,270                 &1.02    &1.27\\
B&	LTT 7987    &2013-07-24 01:19:11      &20:10:56.34   & -30:13:11.2     &115,\,183,\,270                 &1.43    &0.90\\
B&	sky 110 deg &2013-07-24 01:37:15      &14:29:59.43   & +00:00:22.7     &629,\,540,\,600                 &1.83    &1.17\\
B&	sky 90 deg  &2013-07-24 01:56:05      &15:29:53.20   & +14:59:55.6     &629,\,540,\,600                 &1.49    &1.15\\
B&	sky 45 deg  &2013-07-24 02:15:47      &20:59:42.04   & -55:03:00.0     &629,\,540,\,600                 &1.47    &1.24\\
B&	sky 20 deg  &2013-07-24 02:33:39      &20:59:55.00   & -29:59:41.0     &629,\,540,\,600                 &1.31    &1.37\\
B&	sky 13 deg  &2013-07-24 02:51:04      &20:59:55.00   & -21:59:38.0     &629,\,540,\,600                 &1.26    &1.48\\
B&	sky 7 deg   &2013-07-24 03:08:53      &20:59:59.00   & -14:00:00.0     &629,\,540,\,600                 &1.23    &1.59\\
B&	LTT 7987    &2013-07-24 03:26:06      &20:10:56.32   & -30:13:10.5     &115,\,183,\,270                 &1.05    &1.52\\
C&	LTT 3218    &2013-04-19 00:20:50      &08:41:31.35   & -32:56:12.5     &115,\,183,\,270                 &1.03    &1.02\\
C&	LTT 3218    &2013-04-19 00:31:09      &08:41:31.35   & -32:56:12.5     &\,\,\,90,\,\,\,90,\,\,140                   &1.03    &1.40\\
C&	sky 7 deg   &2013-04-19 00:52:32      &08:55:03.20   & +12:58:53.0     &629,\,540,\,600                 &1.32    &1.34\\
C&	sky 13 deg  &2013-04-19 01:09:29      &09:20:04.00   & +12:59:35.0     &629,\,540,\,600                 &1.30    &1.13\\
C&	sky 20 deg  &2013-04-19 01:26:04      &09:45:05.00   & +13:00:36.0     &629,\,540,\,600                 &1.29    &1.28\\
C&	sky 45 deg  &2013-04-19 02:02:04      &11:35:05.00   & +13:00:35.0     &629,\,540,\,600                 &1.27    &1.85\\
C&	sky 90 deg  &2013-04-19 02:19:04      &07:29:47.00   & -75:00:46.0     &629,\,540,\,600                 &1.91    &1.67\\
C&	sky 110 deg &2013-04-19 02:37:31      &17:11:23.00   & -75:03:30.0     &629,\,540,\,600                 &2.29    &1.57\\
C&	LTT 3218    &2013-04-19 02:55:19      &08:41:31.55   & -32:56:14.4     &\,\,\,70,\,\,\,90,\,\,140                   &1.40    &1.39\\
C&	LTT 3218    &2013-04-19 03:04:15      &08:41:31.55   & -32:56:14.4     &115,\,183,\,140                 &1.45    &1.70\\
\noalign{\smallskip}
\hline
\end{tabular}
\tablefoottext{a}{Differential Image Motion Monitor (\url{http://archive.eso.org/wdb/help/eso/ambient_paranal.html})}
\tablefoottext{b}{As noted in the observing logs, these observations had thin cirrus clouds \,\,\,\,\,\,\,\,\,\,\,\,\,\,\,\,\,\,\,\,\,\,\,\,\,\,\,\,\,\,\,\,\,\,\,\,\,\,\,\,\,\,\,\,\,\,\,\,\,\,\,\,\,\,\,\,\,\, \,\,\,\,\,\,\,\,\,\,\,\,\,\,\,\,\,\,\,\,\,\,\,\,\,\,\,\,\,\,\,\,\,\,\,\,\,\,\,\,\,\,\,\,\,\,\,\,\,\,\,\,\,\,\,\,\,\,\,\,\,\,\,\,}
\end{table*}

\section{Analysis}

In this section we describe the current sky background model, focusing on the scattered moonlight component.  Then we discuss how we modified the current scattered moonlight model to improve the aerosol scattering calculations and create a set of models with varying aerosols distributions.  This set of models can then be used to compare with the X-Shooter data to find the most likely aerosol distribution for the different observing runs.

\subsection{Current sky background model}

As part of the Austrian contribution to ESO, the University of Innsbruck In-kind group developed a sky background model.  The model was designed to predict the amount of sky background flux to improve the ETC, used to estimate how long an exposure of a given astronomical target should be for a desired signal-to-noise ratio.  This sky background model produces both an emission and transmission spectrum for a given set of input parameters.  The input parameters describe the conditions of the atmosphere at the time of observation and the geometry of the target and sources of sky background light.  Fig. \ref{eg_spec} shows an example of the sky background model, the scattered moonlight model, the components of the sky background model other than the moonlight, and the transmission curve.  For this case, the model over-predicts the flux between about 0.5 and 1.1\,$\mu$m.  For a full description of the sky background model see \citet{2012A&A...543A..92N}, and for the scattered moonlight model see \citet{2013A&A...560A..91J}.

The sky background model was developed to be used for any instrument at the VLT and the future site of the ELT and can span from 0.3 to 30\,$\mu$m.  It can be decomposed into models for the various sky background sources.

The current scattered moonlight model can be described by the following equation,

\begin{equation}
I_\mathrm{SM}=I_\mathrm{sol}(\lambda)A(\alpha,\lambda)\frac{\Omega_\mathrm{moon}}{\pi}\left(\frac{1}{d_\mathrm{moon}}\right)^2 \xi(z_\mathrm{tar},z_\mathrm{moon},\rho,\zeta,\lambda)F.
\end{equation}

Here, $I_\mathrm{SM} $ is the light intensity from scattered moonlight at the telescope, $I_\mathrm{sol}$ is the solar spectrum, and $A$ is the interpolated albedo of the moon that depends on the lunar phase parameterized by $\alpha$ and wavelength $\lambda$.  $\Omega_\mathrm{moon}$ is the angular size of the moon at the mean distance of the moon from the Earth of 384\,400\,km and equals $6.4177\times10^{-5}$\,sr.  $d_\mathrm{moon}$ is the relative distance between the Earth and moon with respect to the mean distance.  The function that describes the scattering and absorption from molecules and aerosol particles $\xi$ depends on the zenith distance of the target $z_\mathrm{tar}$ and moon $z_\mathrm{moon}$, angular distance between the target and moon $\rho$, properties of the ground and atmosphere $\zeta$, e.g. abundance profiles and ground reflectance, and wavelength $\lambda$.  Lastly is the enhancement factor $F$.  The original scattered moonlight model presented by \citet{2013A&A...560A..91J} had $F=1.2$ to better match the FORS1 data and was used in the ESO sky background model for skycalc and the ETC.  Because of this study, in August 2018, this factor was changed in the ESO sky background model from $F=1.2$ to $F=1.0$.  This change occurred in ESO skycalc version 2.0.4 and the source code for the ESO sky background model release version 1.0.0 as well as in the ETC.  In this paper most of the current models shown for comparisons have $F=1.0$, and when $F=1.2$ the model is labeled model\_1.2.  

With the X-Shooter dataset, we can determine the amount of aerosol extinction from the UV to NIR and test the current sky background model aerosol extinction curve.  Additionally, we can further verify the scattered moonlight model in the optical (where the moonlight is dominate), as well as the full sky background model from the UV to the NIR (0.3 to 2.1\,$\mu$m).  For more details about the sky background model and how it has been tested, see \citet{2012A&A...543A..92N,2013A&A...560A..91J,2014A&A...567A..25N} or the User Manual \citep{skymodel_manual}.

For the scattered moonlight model most of the components (e.g. solar flux, lunar albedo) had measured data that already spanned from the UV to NIR.  The one exception was the aerosol scattering which had only been determined in the optical regime.  The aerosol extinction curve for the VLT was taken from \citet{2011A&A...527A..91P} which was found for the wavelength range 0.4-0.8\,$\mu$m, where an expression based on the best fit values shown in their Fig. 3 was used, namely $\tau_\mathrm{aer}=0.013\times(\lambda_0/\lambda)^{1.38}$ with $\lambda_0=1.0\,\mu$m.  The aerosol optical depth $\tau_\mathrm{aer}$ is related to the extinction coefficient $k_\mathrm{aer}$ by $e^{-\tau}=10^{-0.4k}$.  The aerosol phase function was calculated by mixing different amounts of background aerosols, with the Mie approximation, and finding the phase function where the extinction curve matched the one from  \citet{2011A&A...527A..91P}.  For more details, see \citet{2013A&A...560A..91J}.  The current sky background model uses an extrapolation of the expression from \citet{2011A&A...527A..91P} to longer wavelengths.  Since the aerosol extinction curve has been shown to flatten in the UV, the extinction curve is constant when $\lambda\leq0.4\,\mu$m.  The aerosol phase function was also extrapolated to the full range of the sky model out to 30\,$\mu$m.  It is important to note that the aerosol extinction becomes negligible at redder wavelengths, especially at $\lambda>2.5\,\mu$m.

\subsection{Measuring aerosol extinction}

With the X-Shooter dataset, we can measure the amount of aerosol extinction from 0.3 to 2.1\,$\mu$m and compare this with the current model.  The amount of aerosols present in the atmosphere is not constant and can vary over short and long timescales.  Hence, we will measure the amount of aerosol extinction for the three observing runs separately and compare all three measurements with the current model and each other.

To determine the aerosol extinction curve, we first performed the Langley method using the spectrophotometric standard stars.  At the time X-Shooter was not correcting for ADR.  This caused some of the flux, as a function of wavelength, of the star to be outside of the slit.  We made several attempts to correct for the lost flux (see Section 2 for more details).  Additionally, the standard stars were observed several hours apart for each other.  In this time the atmospheric conditions could have changed.  Because of the flux corrections and the timing of the observations, we found that using these stars for determining the aerosol extinction curve was not accurate enough.  Even an uncertainty of 1\% caused large uncertainties in the aerosol distributions.  This may seem to be in contrast to \citet{2010ApJ...720..811B}, but they were only interested in the overall extinction curve and not in determining the aerosol composition for deriving the phase function.

 Instead of using the spectrophotometric standard stars, we created a set of sky background models, where each model had a different aerosol composition, which we call our aerosol grid.  We then compared each of the sky background models with the observations to determine the best fitting model and most likely aerosol distributions.  We will describe the aerosol grid in more detail below.

The scattered moonlight model has several options for the level of sophistication for the scattering calculations, which vary the number of path and solid angle elements used in these calculations.  The number of elements were found by calculating the amount of scattering for a large range of grid points and finding which values give a decent approximation.  The fewer the grid points, the faster the calculation with a decline in accuracy.  We choose an intermediate mode since we ran 155\,520 models for each observing run, which have 18 observations each (three arms times six $\rho$).  The computation time for each model per observation was around 1.5 seconds.  The most accurate mode for each option takes roughly 3 seconds each and the fastest mode is about 0.5 second each.  With this set-up it took about one month to run the models for one observing run.  We checked the effect of using the less accurate mode, and the change was less than 1.5\% (usually less than 1\%) in flux at all wavelengths for all the angular distances, with the largest effect being for the higher $\rho$ at extreme blue end where the flux is dominated by ozone absorption.

The output of the sky background model gives the radiance spectrum for each and all components, and the transmission spectrum.  This allows us to study only the scattered moonlight portion or the full sky background model.  For an example, see Fig. \ref{eg_spec}.  For more details about the sky background model, see \citet{2012A&A...543A..92N,2013A&A...560A..91J}.

\subsection{Aerosol grid}

\begin{figure}[!ht]
   \centering
   \includegraphics[width=0.49\textwidth]{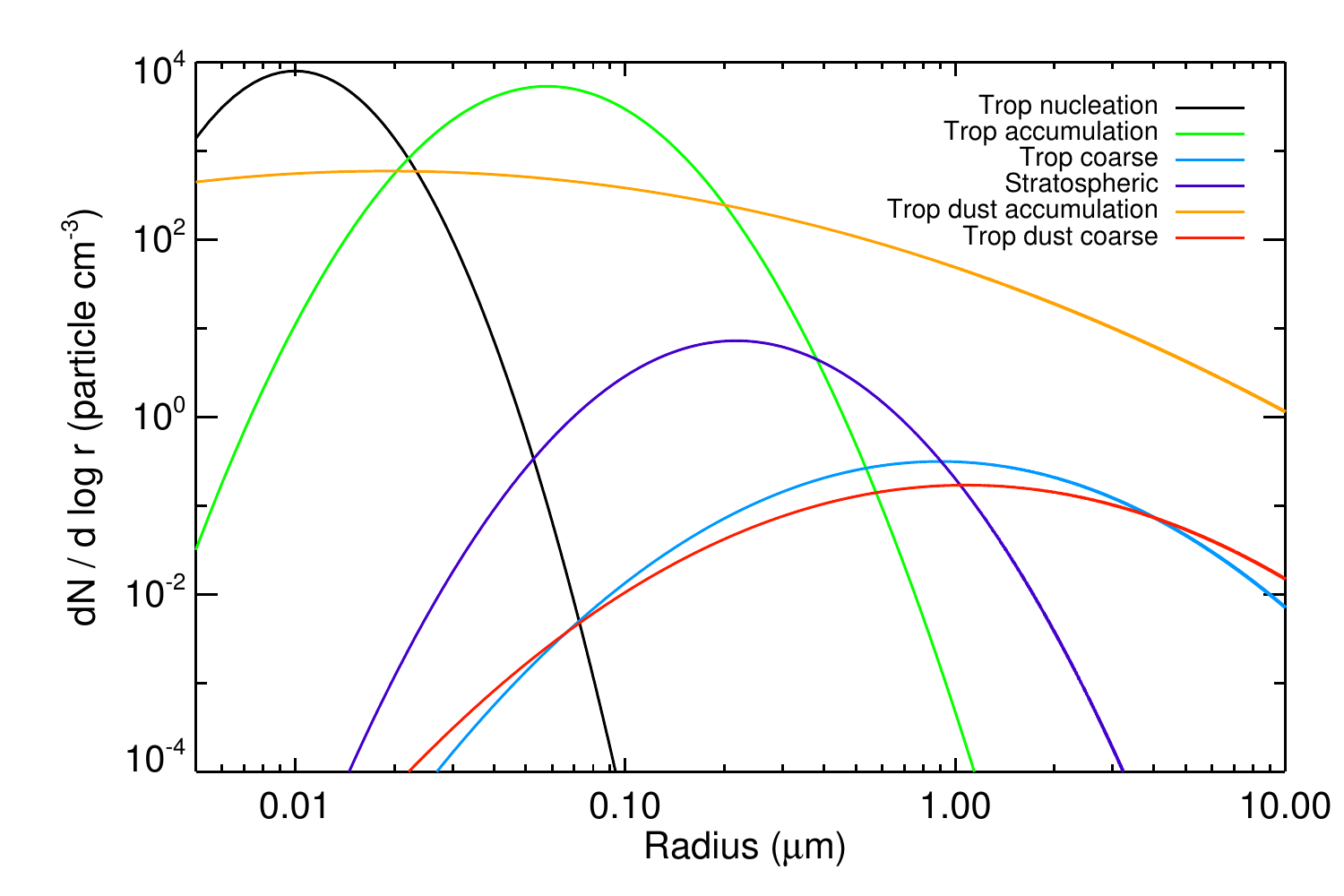}
   \caption{Shows the cumulative number density distribution for the \mbox{different} aerosol modes used as a function of radius for the default \mbox{parameters} given in Table \ref{tab_mie} and Eq. \ref{lognormal}.}
\label{aero_dist}
    \end{figure}

One of the most uncertain parts of the scattered moonlight model is the aerosol scattering because the amount of aerosols can vary on one hour timescales \citep[e.g.][]{2013A&A...549A...8B}.  For the aerosol extinction curve, in the original, optical scattered moonlight model \citep{2013A&A...560A..91J}, we used the one found by \citet{2011A&A...527A..91P}.  We then decomposed this fit into the various aerosol size distributions for remote continental tropospheric and stratospheric modes given by \citet{2012acc.book.....W}.  We scaled the different components to match the aerosol extinction curve and used these scaled components for finding the phase function.  The various aerosol size distributions for the remote continental tropospheric and stratospheric modes are shown in Fig. \ref{aero_dist} and are described by a log normal distribution given by,
\begin{equation}
\frac{dN(r)}{d\log{r}}=\frac{1}{\sqrt{2\pi}}\frac{n}{\log{s}}\exp{\Big[-\frac{(\log{r/R})^2}{2(\log{s})^2}\Big]}.
\label{lognormal}
\end{equation}  
$N$ is the cumulative number density distribution, $n$ is the number density, $R$ is the mean radius of the aerosol particle, and $s$ determines the spread in radii of the particles.  The default parameters for the various aerosols are listed in Table \ref{tab_mie}.  As shown in Fig. \ref{aero_dist}, the tropospheric nucleation mode (black) are only very small particles (less than 0.1\,$\mu$m), while the tropospheric accumulation mode (green) has a broader distribution with particles as large as 1.0\,$\mu$m.  The stratospheric mode (purple) has a similar distribution as the accumulation mode, but shifted to larger particles.  Lastly, out of the remote continental particles, the coarse mode (blue) has the broadest distribution and largest particles of the background aerosols.

Since we do not have a reliable aerosol extinction curve from the spectrophotometric standard stars, we must determine both the aerosol extinction curve and the phase function in another way.  We created a set of sky background models with different amounts of aerosols to compare with all the sky observations for each night.  For each aerosol mode, we calculate an extinction curve and phase function.  For theses calculations, we assume that the aerosol particles are spherical symmetric so we can use the Mie approximation and calculate the Mie extinction curve and phase functions.  For the remote continental aerosols that are described by \citet{2012acc.book.....W}, this is probably a fair assumption \citep[see][for details]{2013A&A...560A..91J}.  For calculating the Mie extinction curve and phase functions we use an IDL code based on \citet{1983asls.book.....B,2004ApOpt..43.5386G}.  These calculations required the distribution of particles (column density, $R$, and $s$), refractive index $N'$, wavelength and $\rho$.  We took $N'=1.40$ for the tropospheric particles and $N'=1.45$ for the stratospheric ones \citep{1966ApOpt...5.1769E,1994JGR....99.3727Y}.  We then took a linear combination of the scaled amounts of each aerosol type to produce the aerosol extinction curve and phase function used in the sky background model.  The resulting extinction curve is a function of wavelength, and the phase function is a function of wavelength and $\rho$.

\begin{figure}[!ht]
   \centering
   \includegraphics[width=0.45\textwidth]{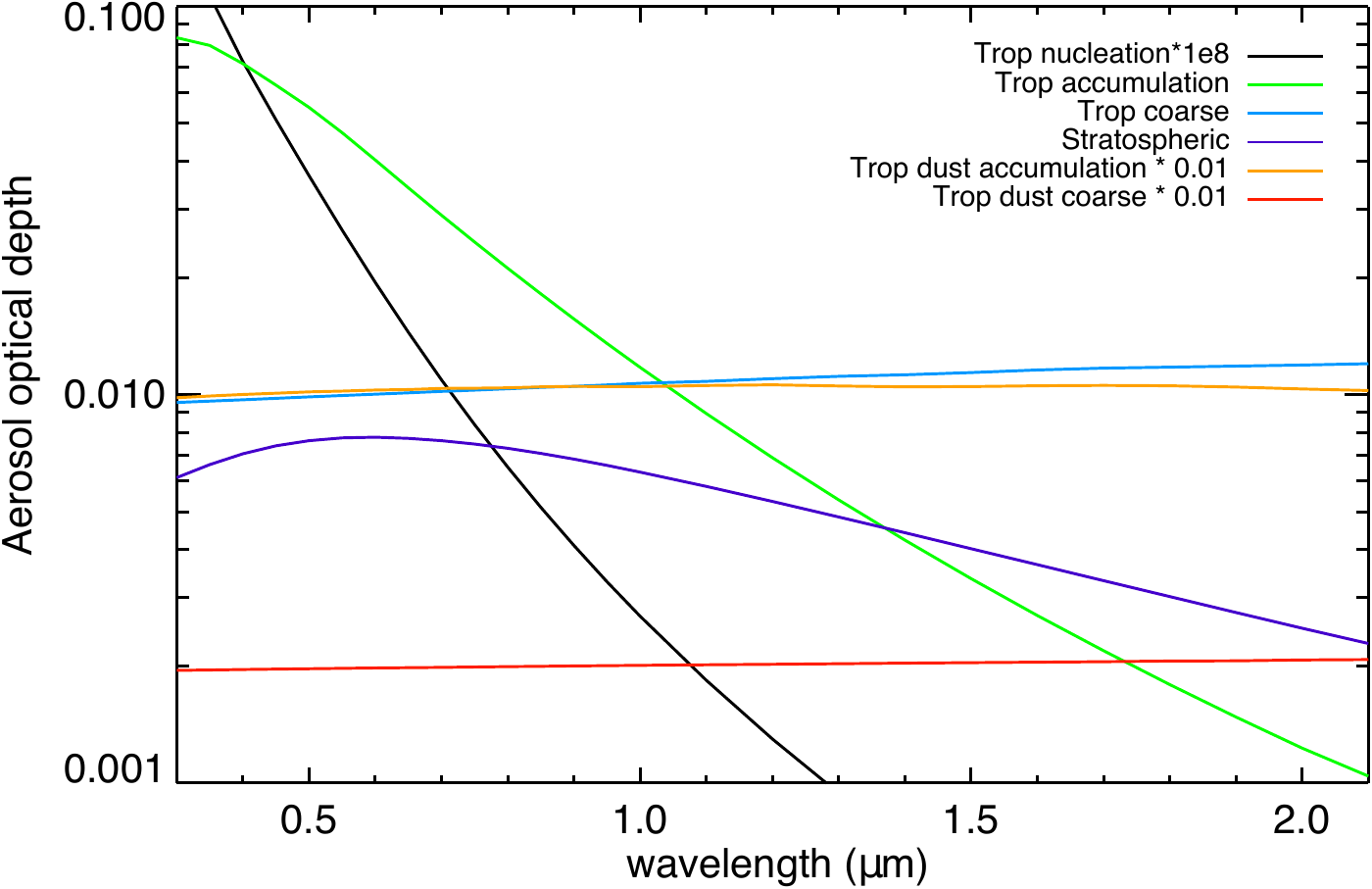}
   \includegraphics[width=0.49\textwidth]{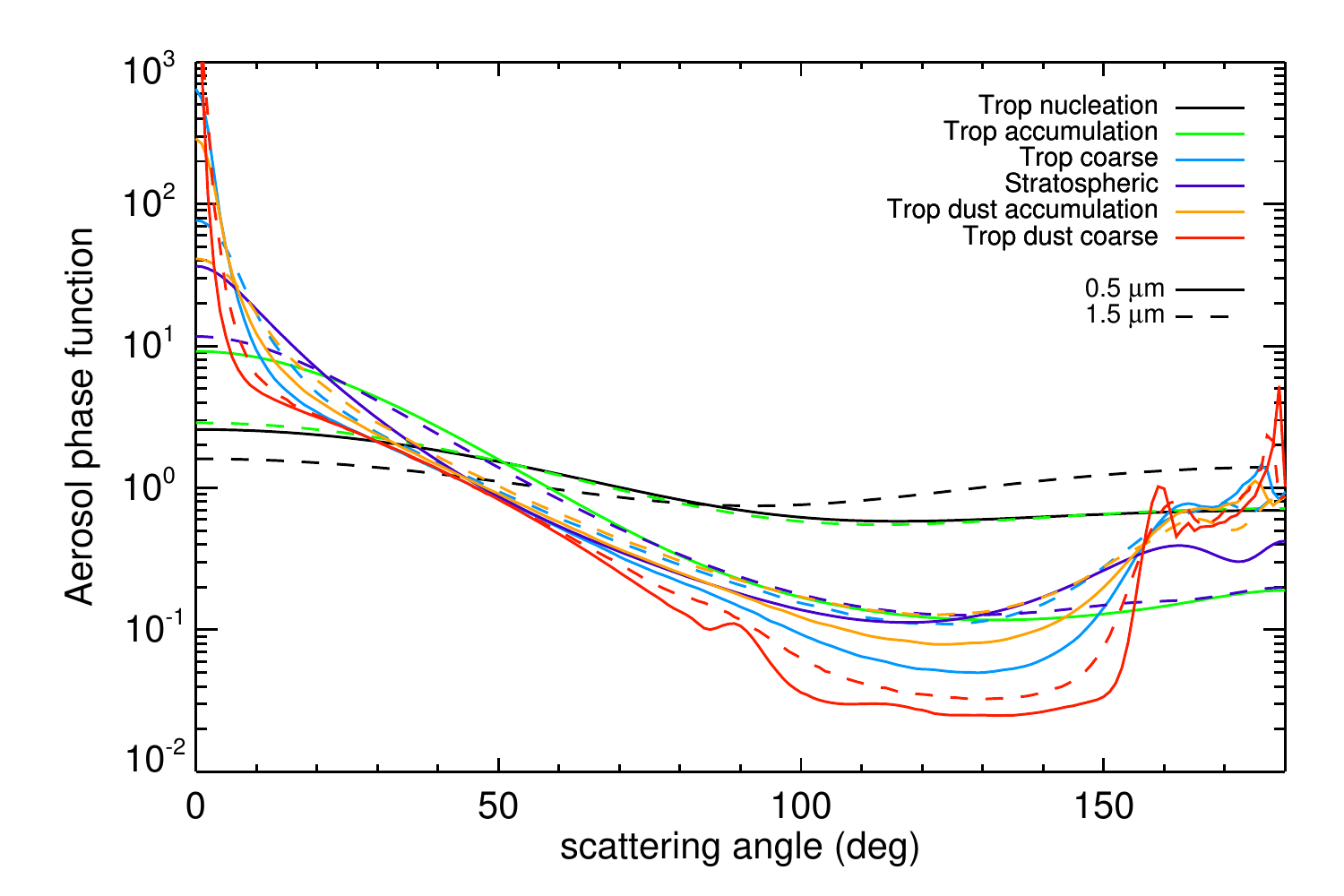}
   \caption{The \textit{top} panel shows the extinction curves for the different aerosol modes used for the wavelength range of X-Shooter (0.3 to 2.1\,$\mu$m) and the \textit{bottom} panel shows the phase function as a function of scattering angle for each aerosol mode used at two different wavelengths, 0.5 (solid) and 1.5 (dashed) $\mu$m.  Both the extinction curves and phase functions were calculated assuming the Mie approximation and were calculated based on the distribution for each aerosol mode with the default values given in Table \ref{tab_mie}.}
\label{aero_exph}
    \end{figure}

We are sensitive to slight changes in the amount of aerosols due to two aspects of this data set. First, our observations cover a wide wavelength range.  Both the aerosol extinction curve and phase function vary with wavelength.  Second, we have observations at multiple distances from the moon taken consecutively to minimize the amount of variation of the atmosphere throughout all the observations.  The phase function is strongly dependent on the angular distance $\rho$ between the moon and target observations (in this case the sky observations).  In Fig. \ref{aero_exph} we show the aerosol extinction curves and phase functions for each type of aerosol particle.  The tropospheric nucleation mode (black) has a negligible contribution to the extinction curve (note it has been enhanced by a factor of $10^8$ so it can be seen on the plot) and additionally has a very flat phase function.  The tropospheric accumulation mode (green) dominates the shape of the extinction curve in the optical, but has a relatively flat phase function.  The tropospheric coarse mode (blue), on the other hand, has a flat extinction curve, but is a very strong function of scattering angle.  The stratospheric mode's (purple) extinction curve is moderately dependent on wavelength as well as the phase function on scattering angle.  The phase functions of the various modes all change slightly as a function of wavelength, shown for 0.5 (solid) and 1.5\;(dashed)\,$\mu$m. 

We produced a regular coarse grid of the different types of aerosols \citep{2012acc.book.....W} that span a reasonable parameter range.  As \citet{2013A&A...560A..91J}, we used the remote continental tropospheric and stratospheric aerosols, since the VLT has a very low amount of aerosols present.  Additionally, we included tropospheric dust aerosols to better fit the $\rho=7^\circ$ and $13^\circ$ observations, which tended to have more flux than what was produced in the sky background model.  The aerosol distribution, extinction curves and phase functions for the dust aerosols are also shown in Figs. \ref{aero_dist} and \ref{aero_exph}, respectively, and listed in Table \ref{tab_mie}.  In Fig. \ref{aero_dist}, the dust distributions are much broader than the other aerosol modes and are larger in $R$ compared to their remote continental counterpart.  The extinction curves for the dust modes in the top panel of Fig. \ref{aero_exph} show that they are fairly flat with respect to wavelength.  Note that the dust extinction curves were reduced to 1\%, since the amount of extinction from the dust aerosols is relatively high compared to the remote continental aerosols.  The phase functions for both are peaked at low scattering angles (shown in the bottom panel of Fig. \ref{aero_exph}), which would increase the flux in the $\rho=7$ and $13^\circ$ observations.  Where the remote continental aerosols are most likely round, for the dust particles, this is less certain.  According to \citet{2002JAtS...59..590D} and references therein, the asymmetry of the dust affects mostly the large scattering angles, $\rho>110^\circ$, beyond our current sky observations.  Thus, for making this calculation feasible we also assumed that they are spherically symmetric and used the Mie approximation.

The default parameters for the different aerosol distributions used are given in Table \ref{tab_mie}.  We then varied seven of these parameters to create our aerosol grid.  The minimum and maximum fraction of each of the default parameters that were varied and the fractional step sizes are shown in Table \ref{tab_mie_var}.  The tropospheric nucleation mode was fixed at the default parameters because this mode has a negligible contribution to the aerosol scattering.  The number density of the tropospheric accumulation mode was varied between 5 and 85\% of the default value.  The shape of the extinction curve in the optical is highly dependent on this parameter as can be seen in the top panel of Fig. \ref{aero_exph}.  The tropospheric coarse mode is important for constraining the phase function, hence we varied all three parameters of this aerosol mode (see bottom panel of Fig. \ref{aero_exph}).  We also varied $n$ of the stratospheric mode.  The stratospheric aerosols  have a different height distribution compared to the tropospheric ones, which nominally starts around 20\,km and with a thickness of about 20\,km \citep{2008AtmRe..90..223D}.  In order to keep the conversion between $n$ and the column density for all aerosol types consistent, we therefore lowered the fraction of $n$ for the stratospheric mode to reflect the difference in effective scale heights to vary between 5 and 55\%.  For convenience we assume that the effective scale heights are 1 km (except for the stratospheric aerosols as discussed above), which is around the actual estimate \citep{1966ApOpt...5.1769E}.  We assume a very small amount of tropospheric dust aerosols and vary the number densities of both between 0.0 and 1.5\%.  This is because we know that the overall amount of aerosols at the VLT is very low, so we must keep the dust aerosols also very low compared to their default value of $n$, otherwise the extinction will be too high.  For reference the current sky background model has 100, 45, 5, and 100\% for the tropospheric nucleation, accumulation, coarse, and stratospheric modes and 0\% for both dust modes, since these were not included in the previous scattered moonlight model.

\begin{table}
\caption{Aerosol modes \label{tab_mie}}
\centering
\begin{tabular}{lllc}
\hline\hline
\noalign{\smallskip}
type & $n$ & $R$ & $\log{s}$\\
 & cm$^{-3}$ & $\mu$m & \\
\noalign{\smallskip}
\hline
\noalign{\smallskip}
Trop nucleation & 3.20 $\times 10^3$ & 0.010 & 0.161  \\
Trop accumulation & 2.90 $\times 10^3$ & 0.058 & 0.217 \\
Trop coarse & 3.00 $\times 10^{-1}$ & 0.900 & 0.380 \\
Stratospheric & 4.49 $\times 10^0$ & 0.217 & 0.248 \\
Trop dust accumulation & 1.14 $\times 10^3$ & 0.019 & 0.770 \\
Trop dust coarse & 1.87 $\times 10^{-1}$ & 1.080 & 0.438 \\
\noalign{\smallskip}
\hline
\end{tabular}
\tablefoot{These values are from \citet{2012acc.book.....W} and trop is an abbreviated form of tropospheric.}
\end{table}

\begin{table}
\caption{Aerosol mode variations \label{tab_mie_var}}
\centering
\begin{tabular}{lclllc}
\hline\hline
\noalign{\smallskip}
type & param & min & max & step & num \\
\noalign{\smallskip}
\hline
\noalign{\smallskip}
Trop accum &	$n$	&        0.05	&0.85	&0.1 & 9\\
Trop coarse &	$n$	&        0.05	&0.85	&0.1 & 9\\
Trop coarse &	$R$	&        0.6    &1.0	&0.1 & 5\\
Trop coarse &	$\log{s}$	&        0.6    &0.9    &0.1 & 4\\
Stratospheric  &	$n$	&        0.05	&0.55	&0.1 & 6 \\
Trop dust accum &	$n$	&        0.000	&0.015	&0.005 & 4\\
Trop dust coarse &	$n$	&        0.000	&0.015	&0.005 & 4\\
\noalign{\smallskip}
\hline
\end{tabular}
\tablefoot{For each aerosol mode (type) and a given parameter in the aerosol grid, this shows the minimum and maximum fraction of the amount as well as the step size of the fraction that was varied.  The fraction is the relative amount of the value for each parameter given in Table \ref{tab_mie}.  The last column (num) provides the number of steps for each parameter, thus the total number of models for each observation is 155\,520=$9\times9\times5\times4\times6\times4\times4$.  Trop and accum are abbreviated forms of tropospheric and accumulation, respectively.}
\end{table}

\section{Verification}

\begin{figure}[!ht]
   \centering
   \includegraphics[width=0.49\textwidth]{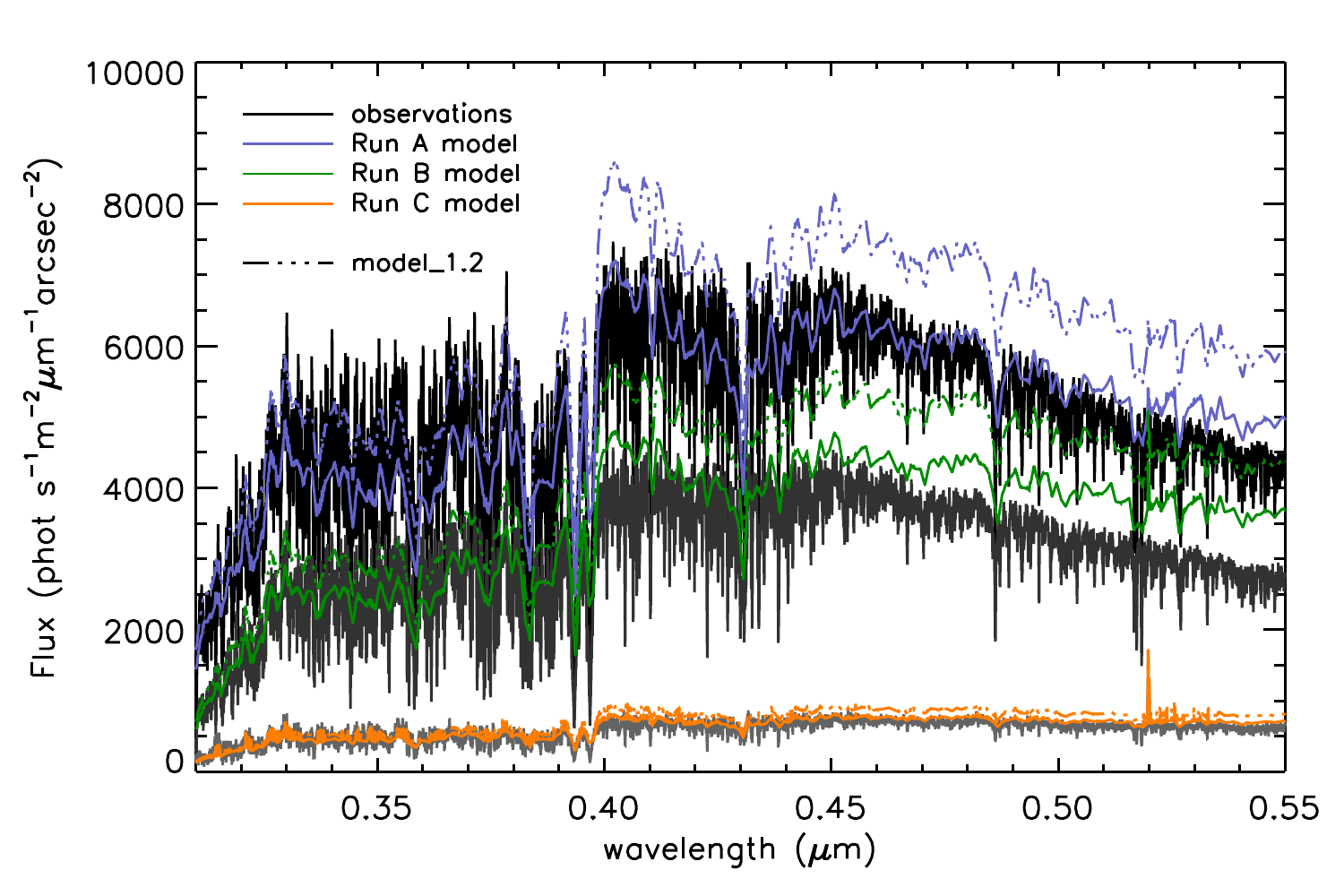}
   \caption{Comparisons between the X-Shooter observations and the current sky background model for different lunar phases (100, 97, 56\%).  Three observations are shown, Run A (black), Run B (dark grey) and Run C (gray), all at $\rho=45^\circ$.  Over-plotted are different models for each run in blue, green and orange for Runs A, B, and C, respectively.  The set of models are the sky background model with $F=1$ (solid) and with $F=1.2$ (dot-dashed).}
\label{ver_lunph}
    \end{figure}
\begin{figure}[!ht]
   \centering
   \includegraphics[width=0.49\textwidth]{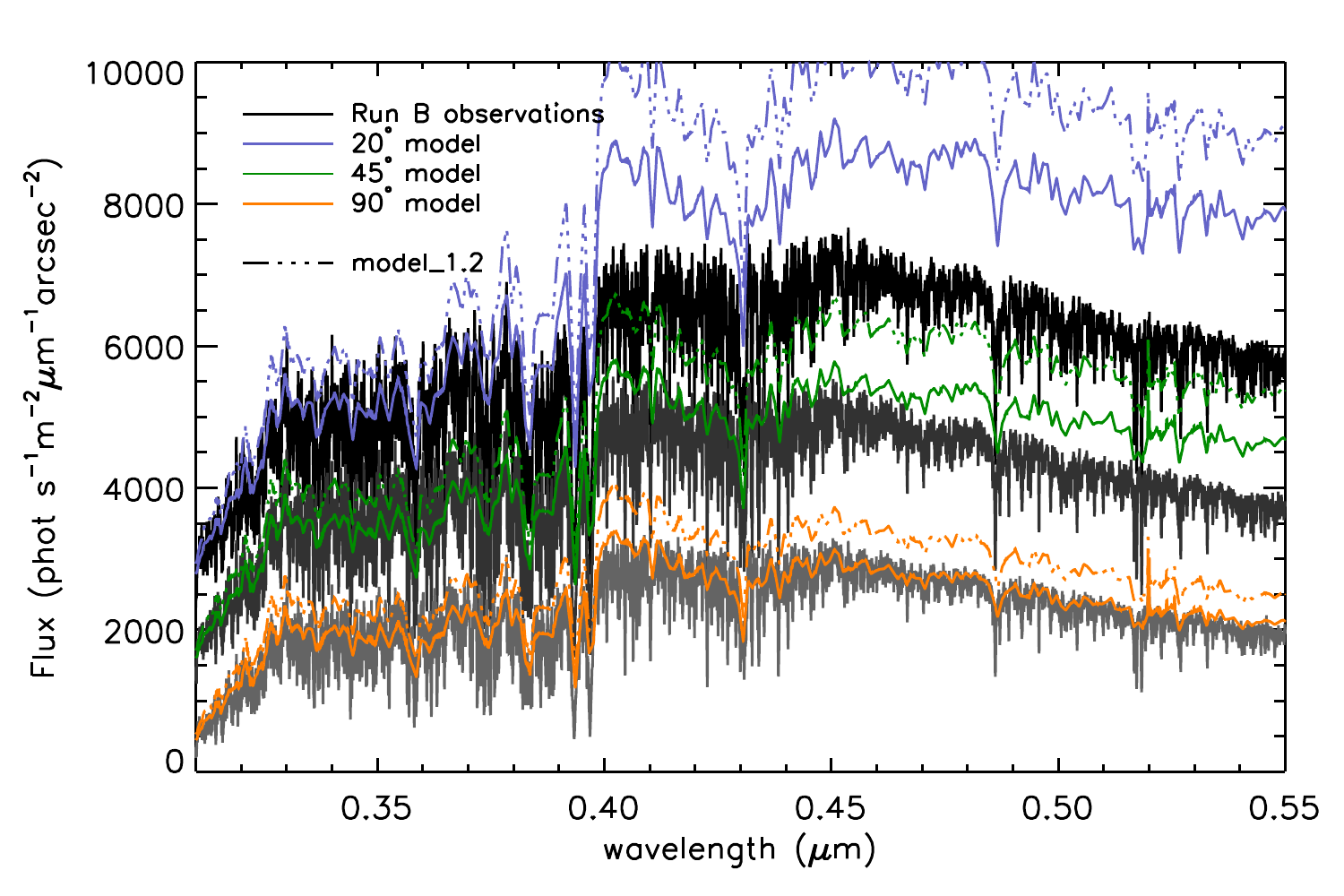}
   \caption{As in Fig. \ref{ver_lunph} but for different angular distances between the moon and target $\rho$.  The three observations shown are from Run B at $\rho=$20 (black), 45 (dark gray), and 90$^\circ$ (gray).  The flux values for $\rho=20^\circ$ were shifted up by 2000 and for $\rho=45^\circ$ by 1000\,phot\,s$^{-1}$m$^{-2}\mu$m$^{-1}$arcsec$^{-2}$ for clarity.}
\label{ver_rho}
    \end{figure}

This set of X-Shooter observations enables us to compare and validate the current scattered moonlight and full sky background models on the basis of three nights with presumably different atmospheric conditions.

The section of the spectrum that is predominately scattered moonlight, but is dominated by Rayleigh instead of aerosol scattering is at the very blue end of the spectra, since Rayleigh scattering depends on $\lambda^{-4}$ and aerosol scattering on $\sim\lambda^{-1}$.  This is where we can compare the current scattered moonlight model with the observations with less uncertainty from the aerosol scattering component.  The small scattering angles $\rho$, even at the blue part of the spectrum are influenced by the aerosol phase function, which is very steep at small $\rho$ (see \textit{bottom} panel of Fig. \ref{aero_exph}).

The original scattered moonlight model was multiplied by a factor $F$ of 1.2 in order to better match the FORS1 observations \citep{2013A&A...560A..91J}.  The reason for needing such a factor was unknown and it was unclear if this was unique to the FORS1 data or a needed factor in general.  With the X-Shooter observations we can conclude that this factor is not needed for these X-Shooter sky observations, so $F$ should be 1.0 and from this study was changed in the ESO sky background model in August 2018.  In Fig. \ref{ver_lunph} we have shown observations for all three runs at $\rho=45^\circ$.  Over-plotted is the sky background model with $F=1.0$ (model) and with $F=1.2$ (model\_1.2).  It is clear that model\_1.2  consistently has too much flux compared with the observations, whereas the sky model with $F=1.0$ is a better match to the data.  The majority of the flux in the plotted wavelength range is coming from scattered moonlight.  The median amount of scattered moonlight flux compared to the total sky background flux in the UVB arm for all three runs, is  96, 95, and 72 \% for Runs A, B, and C, respectively.  The current scattered moonlight model reflects the changes in the lunar phase well.  The effect from the aerosols can be seen in the redder wavelengths shown in Fig. \ref{ver_lunph}.  The effect in Run B is a bit stronger since there seems to be much less aerosols present that night compared to an average night.  There is also a mismatch near $\lambda\sim0.4\,\mu$m, and the cause of this is unclear.

In Fig. \ref{ver_rho}, we made similar comparisons between the observations and models as in Fig. \ref{ver_lunph} but this time looking at the dependence on the angular distance $\rho$.  Three observations from Run B are shown at $\rho=20,\;45,\;\mathrm{and}\;90^\circ$.  Again, the sky background model with $F=1$ is more consistent with the data than the model with $F=1.2$.  The scattered moonlight also dominates at all three $\rho$ in this wavelength range, with the median amount of scattered moonlight in the UVB arm contributing 97, 95, and 93\% of the total background flux for $\rho=20,\;45,\;\mathrm{and}\;90^\circ$, respectively.  The models fit the data well at $\lambda<0.4\,\mu$m and reflect the changes in $\rho$.  The greater mismatch in $\rho=20^\circ$ seems to be mostly from the choice in aerosols (see Fig. \ref{mod_pat} for improved model fits with different aerosol distributions).

For a comparison of all three runs and $\rho=13,\;20,\;45,\;90,\;\mathrm{and}\;110^\circ$ between the current model and the observations, see Fig. \ref{mod_pat}.  In general, the current sky background model is in decent agreement with the data, with the exception of Run A at $\rho=90$ and 110$^\circ$ when there were cirrus clouds (which clouds are not included in the model).  At bluer wavelengths ($\lambda<0.4\,\mu$m), where the scattered moonlight is high and dominated by Rayleigh scattering, the current model is in excellent agreement with the observations.  Additionally, for Run B and C $\rho=90$ and 110$^\circ$, which are much less sensitive to aerosol scattering compared to smaller $\rho$ and dominated by scattered moonlight in the plotted wavelength range, the current model is also in good agreement with the data, as can be seen in Fig \ref{mod_pat}.
 
We can test the other components of the sky background model as well by looking at the observations at redder wavelengths, where the scattered moonlight is a smaller portion of the sky background flux.  Since in this paper we are mostly concerned with the continuum, we will also compare the continuum flux from the sky background model with the observations.  However, the airglow emission, especially at the redder wavelengths, is a critical part of the model and is a very bright component \citep[see e.g.][]{2015ACP....15.3647N}.   For looking at the redder wavelengths, Fig. \ref{eg_spec} shows Run B at $\rho=45^\circ$ with the sky background model for the full wavelength range of X-Shooter.  In this example, the other components of the sky background model (i.e. not scattered moonlight) are shown in orange.  There is a bump around $\lambda=1.0\,\mu$m mostly from the airglow continuum.  In this case, the airglow continuum is overestimated compared to the observations.  Otherwise the sky background model continuum is in good agreement with the data.  

\section{Results}

We ran a set of 155\,520 sky background models with different aerosol compositions for each observing run and observation with the goal to find which models fit the observational data the best.  Since the atmosphere, including the aerosol distributions, can vary nightly (even hourly), we found the best models for each run separately.  First, we calculate the reduced $\chi^2$ for each model for each observation.  This also gives us the best individual model fit for each individual observation.  Then, using the reduced $\chi^2$ for each model of each observation, we determine the most likely aerosol distribution when considering all the observations (except $\rho=7^\circ$).  Additionally, we compute the most likely extinction curve and phase function for each observing run and then run a new model using these.  We also compare the observations with the default, or current, sky background model (as a continuation of Section 4).  The default sky background model is a model with the parametrized aerosol extinction curve, described by \citet{2012A&A...543A..92N} and the phase function determined by \citet{2013A&A...560A..91J}.  We then discuss in more detail the $\rho=7^\circ$ observations.

\subsection{Best individual model fit}

We must quantitatively compare each of the 155\,520 models to each observation.  We want to test the scattered moonlight component and in particular the aerosol scattering, which mainly affects the continuum.  Since we are mostly concerned about how well the continuum of the models matches the observations, we compared them only in certain wavelength regions that were dominated by continuum, i.e. which did not have any strong airglow emission lines or telluric absorption.  We found these so-called \textit{inclusion regions} throughout the X-Shooter spectrum, in all three arms, and tried to have them as evenly distributed throughout the spectrum as possible.  This gets difficult in the redder parts of the spectrum where the sky is dominated by sky lines.  The inclusion regions are marked in Fig. \ref{eg_spec} as pink filled circles.  Then for each spectral arm of X-Shooter, we grouped the inclusion regions into four groups $i$ so arms with more pixels that are inclusion regions (i.e. UVB) do not bias the results.  With these inclusion region groups $i$, we calculated the reduced $\chi^2$ given by,
\begin{equation}
\chi_{M,\rho}^2=\frac{1}{\nu}\sum_{i=1}^4\frac{(O_{i,\rho}-M_{i,\rho})^2}{\sigma_{i,\rho}^2}.
\end{equation}
Here, $M$ stands for the modeled spectrum and $O$ is the observed spectrum.  This is done for each model $M$ and $\rho$.  The parameter $\nu$ is the degrees of freedom derived by number of data points minus the number of free parameters, e.g. $\nu=5\times4\times3-7=53$ for five\;$\rho$, four inclusion groups, three arms, and seven varied aerosol parameters.  Since $\nu$ is not well defined, we re-ran the analysis with the extreme case where $\nu=1$.  The change in the results was minimal (less than 0.5\,$\sigma$) compared to the $\nu=53$ case, and hence the value of $\nu$ is not critical for our analysis.

The uncertainty $\sigma_{\rho}$ is a combination of error from the observations $\sigma_{\rho,O}$ as well as the inaccuracies from the other sky background model components $\sigma_{\rho,M}$.  The error from the observations is dominated by the uncertainty in the flux calibration, namely in the response curves.  The error in the response curve is then multiplied by the observed flux to find $\sigma_{\rho,O}$.  For details about the response curves used for the flux calibrations, see \citet{2015ACP....15.3647N}.  Typical values for $\sigma_{\rho,O}$ relative to $O_\rho-M_\rho$ are five to ten percent for all three arms.  For the uncertainties associated with the model, we are only considering the other model components since we are evaluating the scattered moonlight portion of the model.  The most uncertain component of the continuum comes from the airglow continuum emission.  Based on \citet{2012A&A...543A..92N,2013A&A...560A..91J}, the other model components are accurate to about 20\% for all three arms.  The NIR arm is the most uncertain and we tested the calculations with the assumed 20\%  as well as 50\% $\sigma_{\rho,M}$ in the NIR arm, and the change in the results was negligible.  Then $\sigma_\rho^2=\sigma_{\rho,O}^2+\sigma_{\rho,M}^2$ for each inclusion region group of each arm.

\begin{figure}[!ht]
   \centering
   \includegraphics[width=0.49\textwidth]{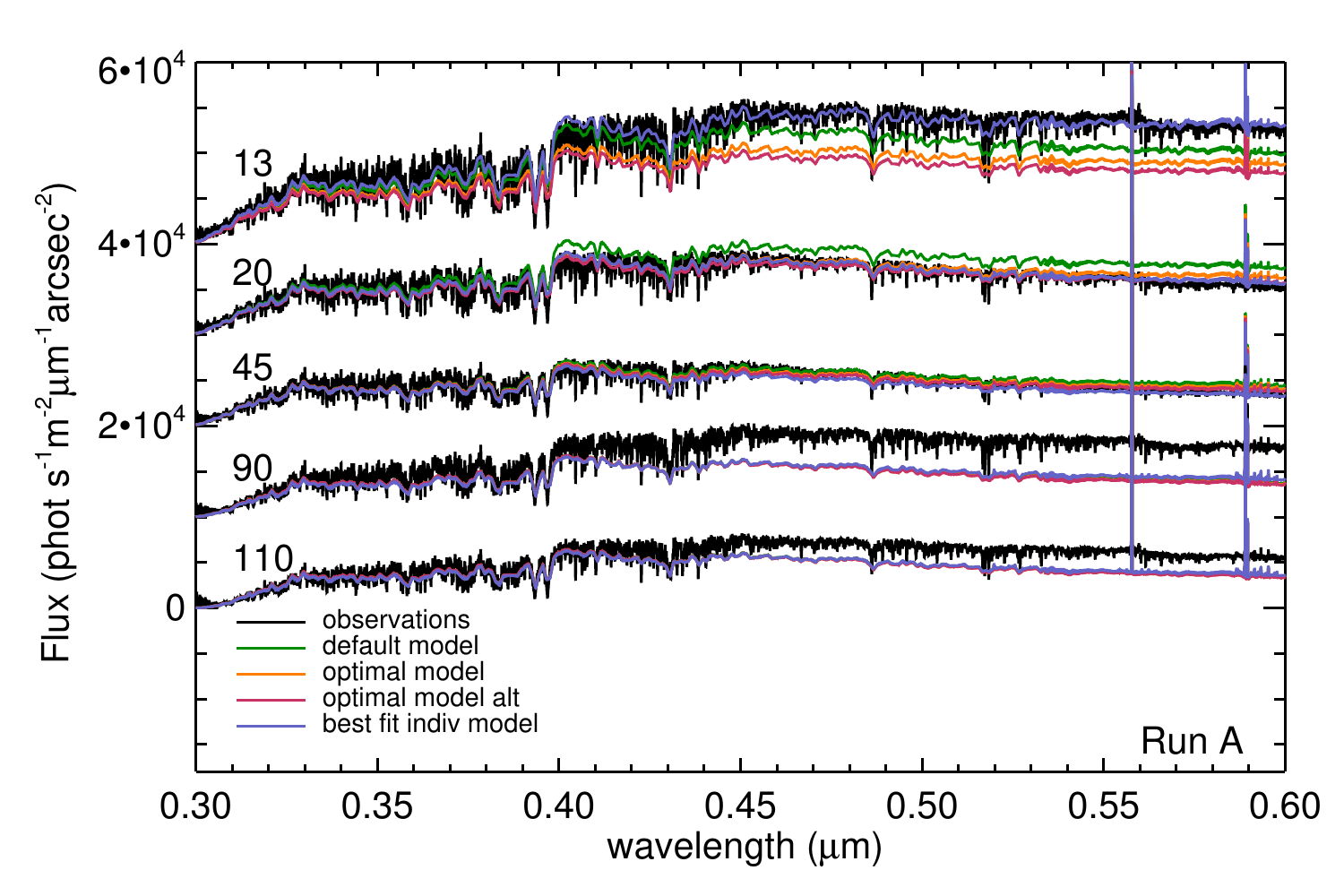}
  \includegraphics[width=0.49\textwidth]{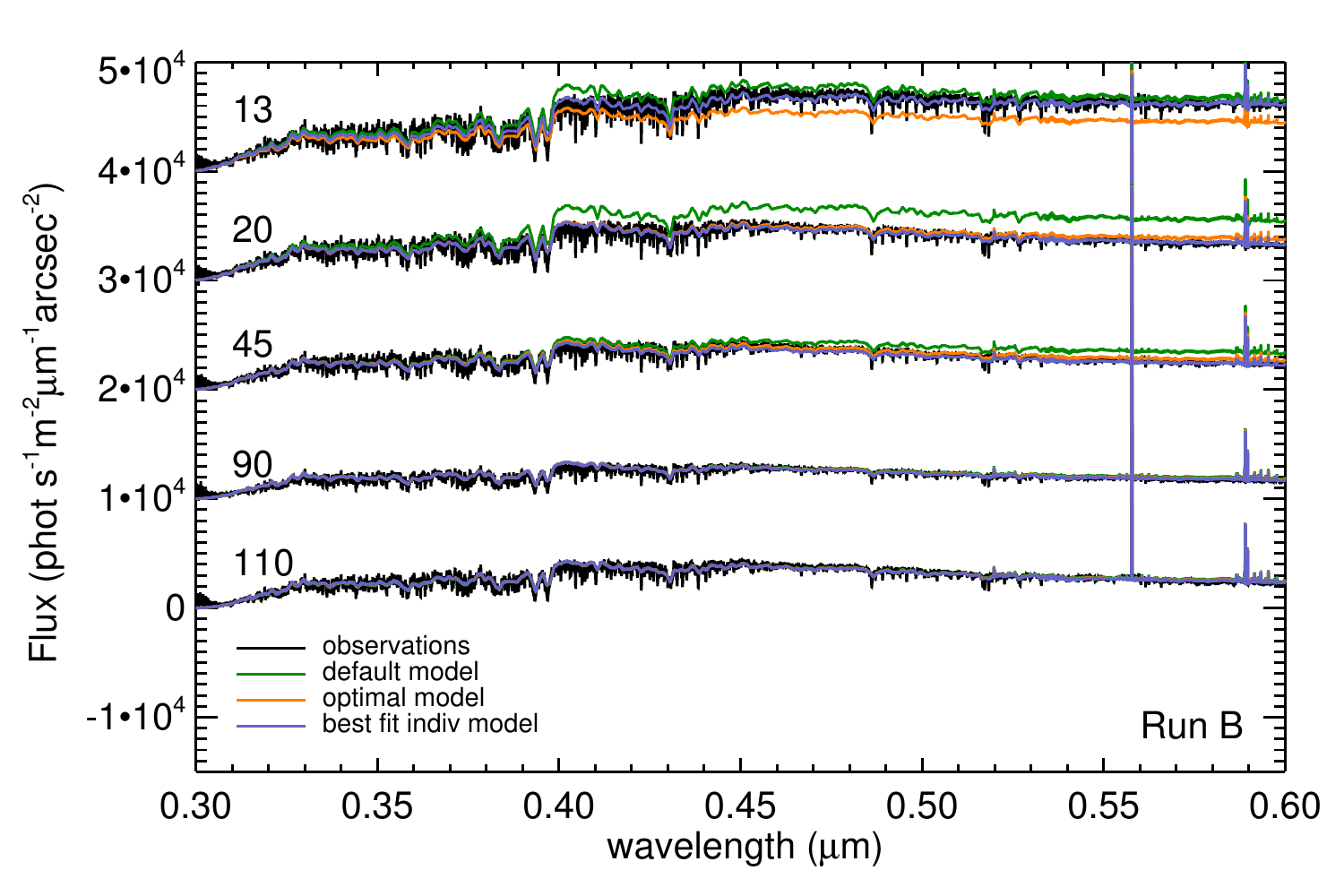}
  \includegraphics[width=0.49\textwidth]{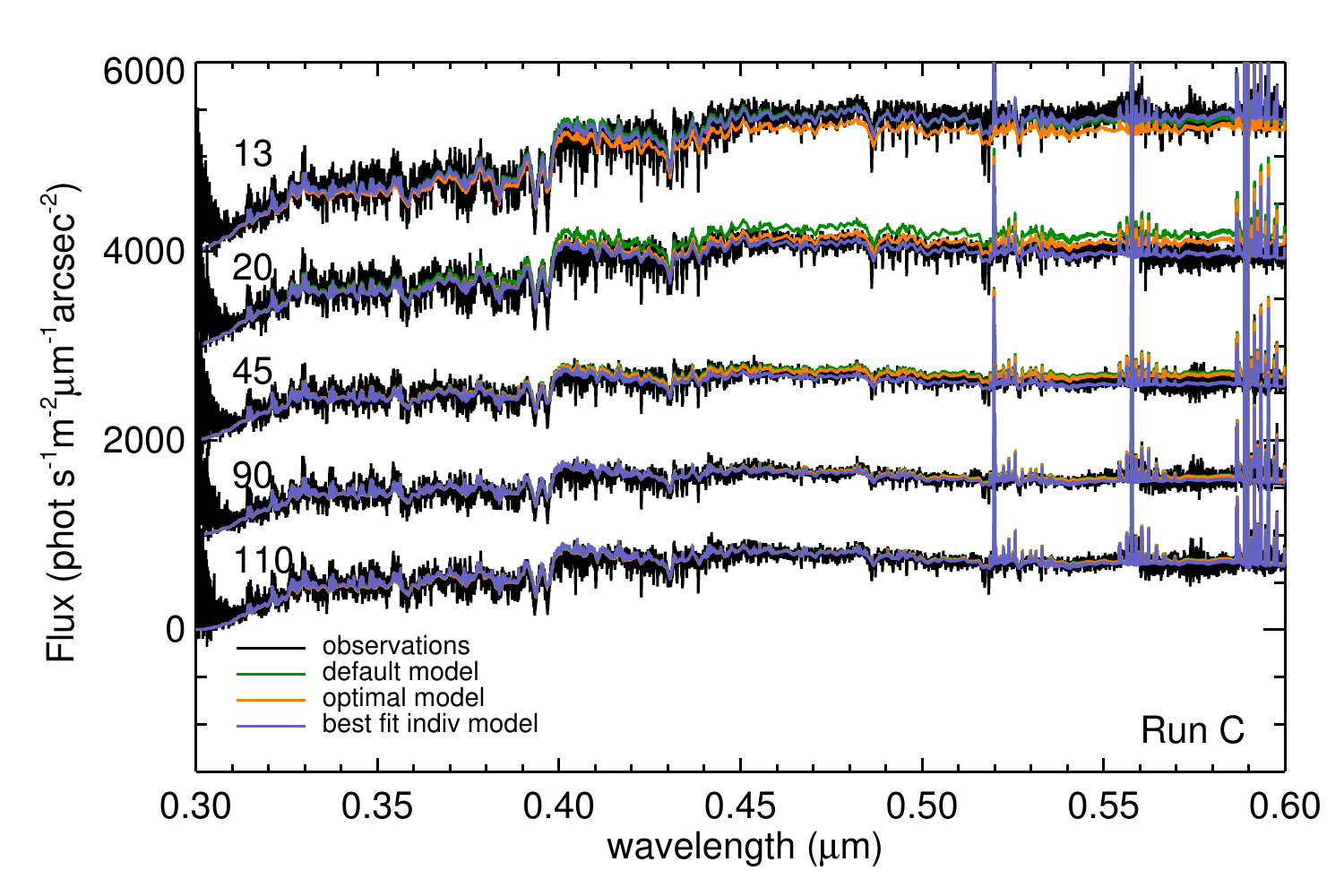}
   \caption{Sky observations (black) with various sky background models over-plotted for Run A (\textit{top}), Run B (\textit{middle}) and Run C (\textit{bottom}).  The observations taken at different $\rho$, 13, 20, 45, 90, and 110$^\circ$, have been offset in flux for clarity. The different models are the current sky background model with $F=1.0$ (green), the model run with the most likely extinction curve (optimal model; orange), and the best fit individual model (blue).  }
\label{mod_pat}
    \end{figure}

Then with the reduced $\chi^2$ for each $M$ and $\rho$, we calculated the likelihood $L$, to find which model for a given $\rho$ was the most likely, for each run, given by:
\begin{equation}
L_{M,\rho}=\exp^{-[\chi_{M,\rho}^2-\mathrm{min}(\chi^2_{M,\rho})]/2}.
\end{equation} 
The most likely model then has $L=1.0$, which we simply refer to as \textit{best}, or the \textit{best individual model fit}.  We also re-ran the analysis without the $\mathrm{min}(\chi^2_{M,\rho})$ normalization, and the results were the same, only the distribution of likelihood values was shifted.  These best models for each $\rho$ and for each run are shown in blue in Fig. \ref{mod_pat}, and the observations $O$ are in black.  For most of the observations, the best model is in excellent agreement with the data.  Run A, $\rho=90$ and $110^\circ$ are an exception.  As noted in the observing logs, there were thin cirrus clouds during those observations, and the models assume photometric clear sky conditions.  The $\rho=7^\circ$ observations for all three runs are also not shown.  This is because the flux in these observations were difficult to reproduce and required a much higher amount of aerosols compared to the other observations.  These data were therefore not used for the rest of the analysis.  A further discussion is in Section 5.3. 

The values for the varied aerosol parameters for the best individual model fit for each $\rho$ and run are given in Table \ref{tab_best_mod}.  There are many degeneracies between several of the aerosol modes.  This can be seen in Fig. \ref{aero_dist}, where several of the aerosol particles overlap, for example the dust and background coarse modes.  This can also be see in Fig. \ref{aero_exph} \textit{top}, where many of the extinction curves for the different modes are fairly flat as a function of wavelength.  The $\rho=7^\circ$ observations for all three runs, have consistently the highest amount of aerosols, which is also inconsistent with the aerosol distributions for the other observations.  This leads to the discussion in Section 5.3 about the $7^\circ$ observations.  The $\rho=13^\circ$ best model also tends to have a higher amount of aerosols than the other observations, but not as high as $\rho =7^\circ$.  It is also interesting that the majority of the observations are consistent with the tropospheric remote continental coarse mode to have a skinner distribution of 60\% of what is quoted by \citet{2012acc.book.....W}, but this could be due to the degeneracy between the remote and dust coarse modes. 

\begin{table*}
\caption{Aerosol parameters for the best individual and optimal models \label{tab_best_mod}}
\centering
\begin{tabular}{clccccccc}
\hline\hline
\noalign{\smallskip}
 && Tropospheric & Trop  & Trop & Trop &Stratospheric & Trop & Trop \\
  && accumulation  & \;coarse\; & \;coarse\; & \;coarse\; & & dust accum  & dust coarse   \\
 Run\;\; & $\rho$ & $n$ & $n$ & $R$ & $\log{s}$ & $n$ & $n$ & $n$ \\
\noalign{\smallskip}
\hline
\noalign{\smallskip}
  A & $7^\circ$\tablefootmark{a} & 0.85 & 0.85 & 1.0 & 0.9 & 0.55 & 0.015 & 0.015 \\
  A & $13^\circ$ & 0.55 & 0.65 & 0.6 & 0.6 & 0.55 & 0.010 & 0.015 \\
  A & $20^\circ$ & 0.35 & 0.75 & 0.6 & 0.8 & 0.05 & 0.000 & 0.000 \\
  A & $45^\circ$ & 0.15 & 0.05 & 0.6 & 0.6 & 0.05 & 0.000 & 0.000 \\
  A & $90^\circ$\tablefootmark{b} & 0.85 & 0.85 & 1.0 & 0.9 & 0.55 & 0.015 & 0.015 \\
  A & $110^\circ$\tablefootmark{b} & 0.85 & 0.85 & 1.0 & 0.6 & 0.55 & 0.015 & 0.015 \\
  B & $7^\circ$\tablefootmark{a} & 0.85 & 0.85 & 0.8 & 0.6 & 0.55 & 0.015 & 0.000 \\
  B & $13^\circ$ & 0.25 & 0.25 & 1.0 & 0.9 & 0.55 & 0.005 & 0.015 \\
  B & $20^\circ$ & 0.15 & 0.05 & 0.9 & 0.6 & 0.05 & 0.000 & 0.010 \\
  B & $45^\circ$ & 0.05 & 0.05 & 0.6 & 0.6 & 0.05 & 0.000 & 0.000 \\
  B & $90^\circ$ & 0.05 & 0.05 & 0.6 & 0.6 & 0.05 & 0.000 & 0.000 \\
  B & $110^\circ$ & 0.05 & 0.05 & 0.6 & 0.6 & 0.05 & 0.000 & 0.005 \\
  C & $7^\circ$\tablefootmark{a} & 0.85 & 0.85 & 1.0 & 0.9 & 0.55 & 0.015 & 0.015 \\
  C & $13^\circ$ & 0.45 & 0.85 & 1.0 & 0.8 & 0.55 & 0.000 & 0.005 \\
  C & $20^\circ$ & 0.35 & 0.05 & 0.6 & 0.6 & 0.05 & 0.000 & 0.010 \\
  C & $45^\circ$ & 0.15 & 0.05 & 0.6 & 0.6 & 0.05 & 0.000 & 0.000 \\
  C & $90^\circ$ & 0.05 & 0.05 & 0.6 & 0.6 & 0.05 & 0.000 & 0.000 \\
  C & $110^\circ$ & 0.05 & 0.05 & 0.6 & 0.6 & 0.05 & 0.000 & 0.000 \\
  A  & optimal & 0.27(12) & 0.46(26) & 0.80(14) & 0.74(11) & 0.28(17) & 0.006(4) & 0.006(6) \\
  A  & optimal alt\tablefootmark{c} & 0.22(10) & 0.45(26) & 0.80(14) & 0.74(11) & 0.28(17) & 0.004(3) & 0.006(6) \\
  A  & optimal (no dust) & 0.39(12) & 0.61(22) & 0.85(13) & 0.75(11) & 0.36(16) & $\ldots$ & $\ldots$ \\
  B  & optimal & 0.13(8) & 0.44(26) & 0.80(14) & 0.74(11) & 0.28(17) & 0.003(3) & 0.007(6) \\
  B  & optimal (no dust) & 0.17(9) & 0.54(24) & 0.82(14) & 0.75(11) & 0.35(16) & $\ldots$ & $\ldots$ \\
  C  & optimal & 0.33(18) & 0.44(26) & 0.80(14) & 0.75(11) & 0.29(17) & 0.006(5) & 0.007(6) \\
  C  & optimal (no dust) & 0.43(18) & 0.46(26) & 0.80(14) & 0.75(11) & 0.31(17) & $\ldots$ & $\ldots$ \\
  \noalign{\smallskip}
\hline
\noalign{\smallskip}
\end{tabular}
\tablefoottext{a}{See Section 5.3 about the $7^\circ$ observations}
\tablefoottext{b}{Had thin cirrus clouds}
\tablefoottext{c}{Excluding the 90 and 110$^\circ$ observations\,\,\,\,\,\,\,\,\,\,\,\,\,\,\,\,\,\,\,\,\,\,\,\,\,\,\,\,\,}
\tablefoot{The values for the varied parameters for each observation of the best fit model determined by Eq. 3 and the most likely, optimal values ($<x>$ with $1\,\sigma_x$ in parenthesis) for each Run by Eq. 6}
\end{table*}

\subsection{Most likely models and aerosol size distributions}

We have found the best model for each run and each observation, discussed in the above Section 5.1.  Next we will determine the likelihoods for each model, considering $\rho=13,\;20,\;45,\;90,\;110^\circ$  observations for a given run.  Each run was taken on one given night and the aerosol properties should not change too much during the few hours of observation.  The observations taken at different $\rho$ are each sensitive to different aspects of aerosol scattering process, so by combining them we can find which model reproduces the observations the best for a given run.  

We follow a similar procedure as what was done in Section 5.1, but now Eq. 4 is modified to:
\begin{equation}
L_{M}=\exp^{-[\chi_{M}^2-\mathrm{min}(\chi^2_{M})]/2},
\end{equation} 
where $\chi_M^2=\chi_{13^\circ}^2+\chi_{20^\circ}^2+\chi_{45^\circ}^2+\chi_{90^\circ}^2+\chi_{110^\circ}^2$.  We did not include $\rho=7^\circ$ observations because of higher flux and therefore higher aerosol content in the most likely models (see Section 5.3).  


We can use the sky background models and their likelihoods to find the most likely aerosol properties for each run.  We would expect each night to have slightly different aerosol extinction curves due to the changing atmospheric conditions \citep[see for e.g.][]{2013A&A...549A...8B}.

We followed Eqs. 8 and 9 of \citet{2009A&A...507.1793N} to find the weighted mean $<x>$ and standard deviation $\sigma_x$ of the several different aerosol properties (extinction curve, phase function, varied parameters of the aerosol distribution) using the likelihoods $L$ of each model $i$, shown again here:
\begin{equation}
<x>=\frac{\sum^b_{i=1}L_i x_i}{\sum^b_{i=1}L_i},\quad\quad
\sigma_x=\sqrt{\frac{\sum^b_{i=1}L_i(x_i-<x>)^2}{\sum^b_{i=1}L_i}}.
\end{equation}
The index $i$ runs through all the models for the given run.  $L_i$ is the weight for each model, and is taken from Eq. 5.  

We have calculated the most likely aerosol extinction at each wavelength, so $x=\tau_\mathrm{aer}$ and $\sigma_x=\sigma_\tau$.  The resulting optimal aerosol extinction curves for each run are shown in Table \ref{tab_ext}.  Run A and C were conducted both in April, while Run B was in July.  Run A and C have more similar aerosol extinction curves than Run B as expected.  The observations at 90 and 110$^\circ$ for Run A had thin cirrus clouds.  We additionally calculated the best aerosol extinction curve for Run A without these two observations, called Run A Alternate (Alt).

\begin{figure}[!ht]
   \includegraphics[width=0.49\textwidth]{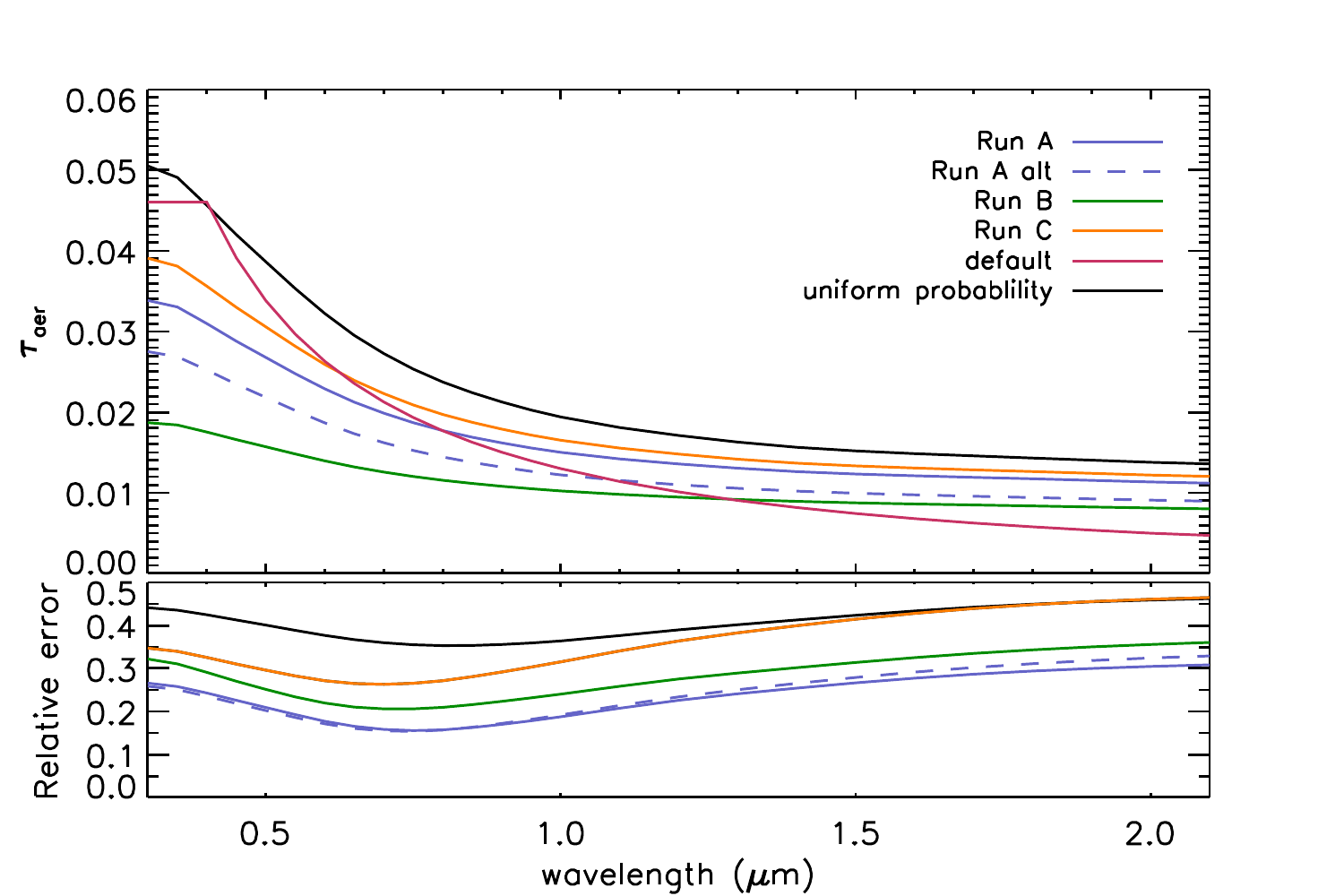}
   \caption{\textit{Top} panel shows the aerosol extinction curves at the zenith for the wavelength range of X-shooter.  These are the most likely, optimal aerosol extinction curves for each run discussed in Section 5.2 .  For Run A, an alternate optimal curve (dashed) is also shown, which is the same calculation but excluding the 90 and 110$^\circ$ observations.  The default curve is the one currently used in the ESO sky background model.  Lastly, the uniform probability curve is derived with Eq. 6 but with all $L_i=1$.  The \textit{bottom} panel shows the relative error (1\;$\sigma_{\tau}/\tau$).  The values and errors for the extinction curves are also given in Table \ref{tab_ext}.}
\label{best_ext}
    \end{figure}

In Fig. \ref{best_ext} we show several different aerosol extinction curves.  The most likely aerosol extinction curves derived using Eq. 6 for each run are shown.  The \textit{top} panel gives the aerosol extinction curves and the \textit{bottom} panel the relative errors ($\sigma_\tau/\tau$).  Additionally, we plotted the default extinction curve used, which is flat at $\lambda\leq0.4\,\mu$m and then follows the curve from \citet{2011A&A...527A..91P}, $\tau=0.013\times(\lambda_0/\lambda)^{1.38}$ where $\lambda_0=1.0\,\mu$m.  The default curve is in somewhat agreement with Run C at short wavelengths, but too high compared to the other runs.  Then in the optical range between 0.5 and 1.0\,$\mu$m it is consistent with Run A and C, and red-wards of about 1.0\,$\mu$m it dips down and stays lower than Run A and C.  Run B, on the other hand, has a lower extinction curve at all wavelengths compared to Run A and C and is much flatter than the default curve.  For completion we over-plotted an extinction curve with uniform probability, so all $L_i=1$.  This curve is steeper than the other curves and usually has larger errors.

In general, the best extinction curves for all three runs are consistent with there being some changes in the aerosol content of the atmosphere both nightly and seasonally.

\begin{figure}[!ht]
   \includegraphics[width=0.49\textwidth]{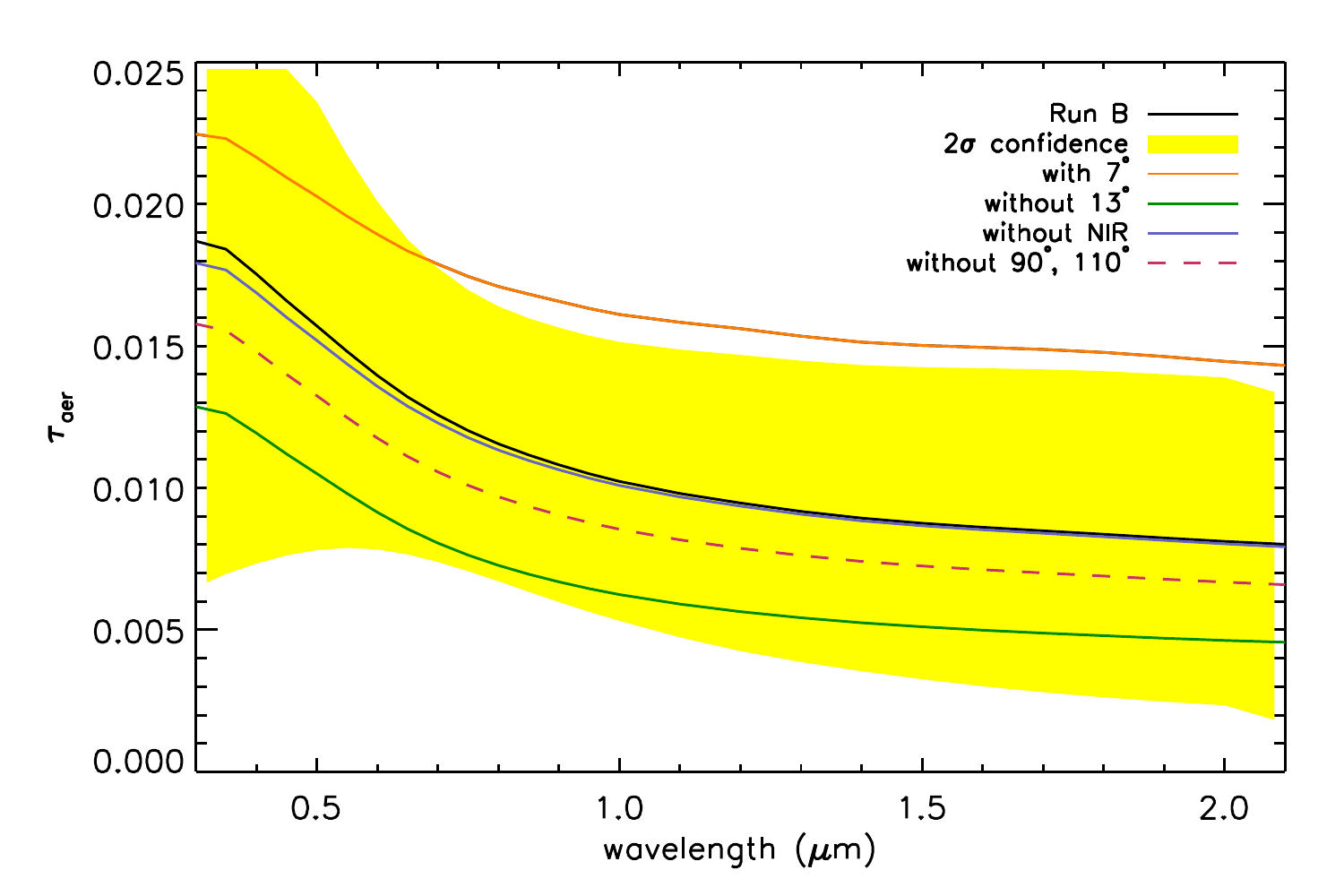}
   \caption{Shows the robustness of the analysis for determining the most likely aerosol extinction curve.  The standard analysis (Run B, black), which includes all observations except $\rho=7^\circ$, with the yellow region giving the 2\,$\sigma$ confidence interval, is plotted.  Over-plotted are the same procedures but including $\rho=7^\circ$ (orange), excluding $\rho=13^\circ$ (green), excluding the NIR arm (blue), and excluding $\rho=90$ and $110^\circ$.}
\label{diff_b}
\end{figure}

We tested the robustness of our optimal extinction curves and how they depended on our choice of $\rho$, inclusion regions, and airglow model.  The resulting extinction curves for Run B are shown in Fig. \ref{diff_b} and the other two runs showed similar results.  The standard analysis included data from all $\rho$ except $\rho=7^\circ$.  When we include data with $\rho=7^\circ$, as expected, the optimal extinction curve is higher at all wavelengths.  It is roughly $2\sigma$ higher than the standard values.  When we exclude both $\rho=7$ and $13^\circ$ data, the extinction curve has lower values, within $1.2\,\sigma$ of the standard.  For Run A, we additionally showed results excluding the $\rho=90$ and $110^\circ$ data due to poor seeing conditions.  Here we show how excluding these data would affect Run B and it lowers the values, but within $1\,\sigma$.  We also tried completely excluding the NIR arm from the analysis.  Since the scattered moonlight contribution in NIR is small, its effect on the results should also be minimal.  Indeed in Fig. \ref{diff_b}, excluding the NIR arm has a negligible effect.  We also re-ran the full analysis multiple times, where each time we excluded one of the inclusion region groups, and this also had a negligible effect on results.  Lastly, we tested if the airglow model affected the results.  The airglow continuum is one of the most uncertain parts of the sky background model because it is difficult to measure and varies on short and long timescales as well as spatially.  In Fig. \ref{eg_spec} the sky background model flux starts to increase around $0.75\,\mu$m due to the other components (orange line).  This is mainly from the airglow continuum model used.  A new template for airglow was created based on another set of X-Shooter data which contained only dark time observations \citep{2015ACP....15.3647N}.  The starlight and zodiacal light components were removed using the sky background model \citep{2012A&A...543A..92N} and then taking the median of the remaining airglow continuum and emission lines.  We scaled this template using the average airmass of the template and the airmass of our sky observations.  This scaled template seemed generally to match our X-Shooter observations better as expected since this was optimized for X-Shooter data.  We re-ran the analysis using this airglow template and the changes were negligible.  In summary, the choice of data from given $\rho$ does affect the results, especially including or excluding the small $\rho$.  However, the optimal aerosol extinction curve is not sensitive to the inclusion regions used or how we estimate the airglow.  
 
\begin{figure}[!ht]
   \includegraphics[width=0.49\textwidth]{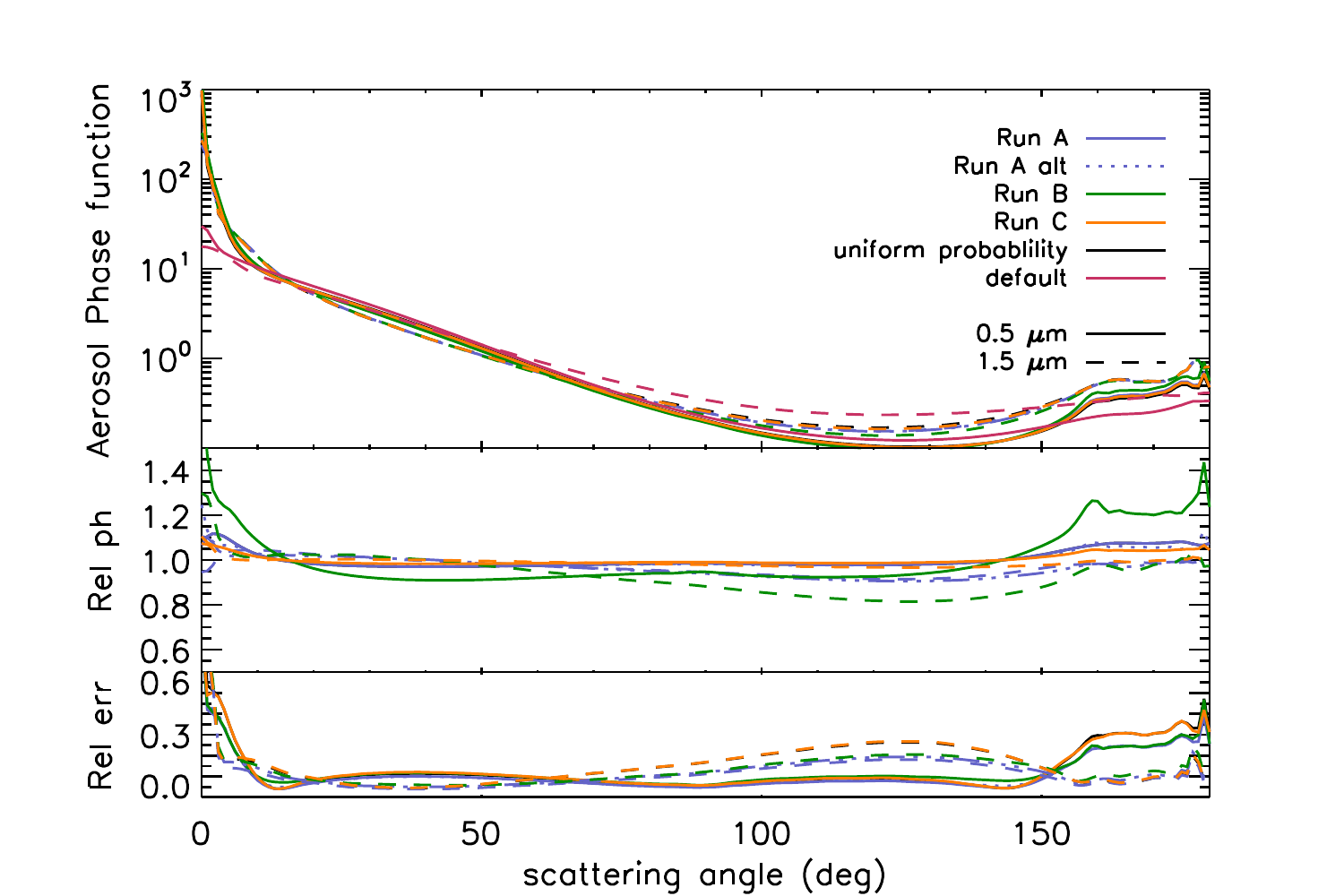}
   \caption{\textit{Top} shows the aerosol phase function as a function of scattering angle for the most likely models at two different wavelengths, 0.5 and 1.5\,$\mu$m.  For Run A, alternate optimal functions (dotted and dashed-dotted) are also shown, which are the same calculations but excluding the 90 and 110$^\circ$ observations for both wavelengths.  Also shown is the phase function used in the current sky background model (default) and with uniform probability $L_i=1$ in Eq. 6.  For clarity, the \textit{middle} panel shows the three runs relative to the uniform probability phase function.  The \textit{bottom} panel shows the relative error $\sigma_x/<x>$ (from Eq. 6).}
\label{best_ph}
    \end{figure}

We also calculated the most likely aerosol phase functions for each run using Eq. 6.  The results are plotted in Fig. \ref{best_ph} for $\lambda=0.5$ and 1.5\,$\mu$m.  As in Fig. \ref{best_ext}, we show the most likely phase functions for each run, the default phase function, and the phase function calculated with all $L_i=1$ called uniform probability for both wavelengths.  The \textit{top} panel shows the phase functions, while the \textit{middle} panel shows the phase functions relative to the one with uniform probability to see more easily any differences amongst the runs.  The \textit{bottom} panel shows the relative error ($\sigma_x/<x>$).  The default phase function is less peaked at low scattering angles and is smoother at high angles compared to others.  As evident in the \textit{middle} panel, Run B has a slightly higher phase function at both low and high scattering angles compared to the other runs.  The phase function for most scattering angles is well constrained (around 10\%) and increases towards the very small and large $\rho$.  Since our lowest usable observation is at 13$^\circ$ and our highest is at 110$^\circ$, this lack of constraint there is not surprising.

With the most likely aerosol extinction curves and phase functions for each run, we ran a new set of models using these to directly compare with the observations.  These new models are referred to as the \textit{optimal} models for each run.  These optimal models are shown in Fig. \ref{mod_pat} in orange.  From the figure, the optimal models tend to underestimate the flux in $\rho=13^\circ$ observations, but otherwise fit the data well.  The optimal models are also generally in better agreement with the observations than the current sky background model.

We can take this a step further and determine the most likely values for the varied parameters in the aerosol distributions using Eq. 6.  The means and uncertainties for each are given in Table \ref{tab_best_mod}.  Due to the degeneracies, many of the parameters are not well constrained, but the amount of tropospheric accumulation mode prefers lower values than the uniform probability provides.  Here, we also show the distribution if we ignore the dust modes, and there are only slight changes. 

\begin{figure}[!ht]
   \includegraphics[width=0.49\textwidth]{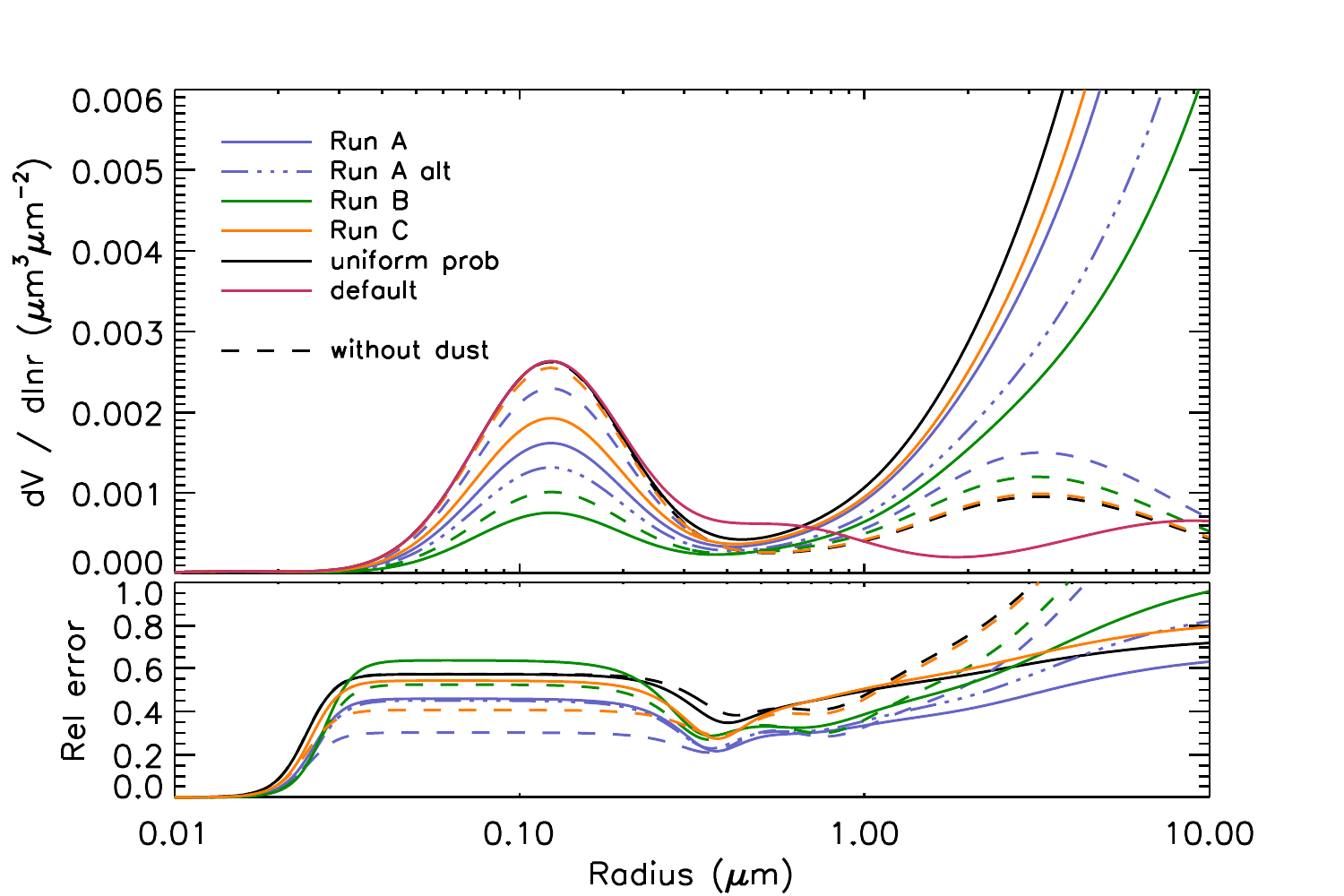}
   \caption{\textit{Top} shows the most likely column-integrated volume densities of aerosol particles as a function of radius for Run A (blue, solid), Run A alternate (blue, dot-dashed), Run B (green) and Run C (orange).  The volume density with uniform probabilities (black) and the distribution used in the current sky background model (default, pink) are also shown.  The same analysis, but without the dust particles is shown for all three runs and with uniform probability (dashed).  The \textit{bottom} panel shows the relative error, given by $\sigma_x/<x>$.}
   \label{best_vden}
    \end{figure}

With the likelihood weights for each model, we can also find the most likely column-integrated volume density of aerosols for each run. In Eq. 6, $x$ now refers to the volume density, which can be calculated from the number density given in Eq. 2, times the volume of the particle at a given radius, and then integrated over the column assuming the 1\,km effective scale height, for all aerosol types. The most likely column-integrated volume density of aerosols as a function of size for all three runs, including Run A alternate, is shown in Fig. \ref{best_vden} along with having uniform probability, the runs without dust, as well as the default distribution from the current sky background model.  This default distribution was taken from \citet{2013A&A...560A..91J} used for calculating the aerosol phase functions.  The \textit{top} panel shows the distributions and the \textit{bottom} panel shows the relative errors.  The volume distributions derived are similar to ones seen in other remote areas with AERONET\footnote{https://aeronet.gsfc.nasa.gov/} \citep{1998RSEnv..66....1H}, e.g.  Mauna Loa Observatory \citep{2001JGR...10612067H}.  AERONET is a network of robotic aerosol detection sites placed all over the world, and the Mauna Loa Observatory is known to have a very low aerosol optical depth, like Cerro Paranal.  In Fig. \ref{best_vden}, there is a peak around $R=0.15\,\mu$m from the accumulation mode, where Run B has the smallest peak alluding to there being a low amount of aerosols present that night compared to the other two runs.  The distributions without dust also tend to have a larger peak at $R=0.15\,\mu$m compared to the distributions from the same runs with dust.  Then there is a second peak at larger radii.  For the distributions without dust this peak is usually lower than the one at smaller radii at around $R=3\,\mu$m.  With dust, the distributions increase steadily and peak at $R>10\,\mu$m.  These bi-modal peaks in the volume distribution have also been seen in other remote AERONET sites, and seem to be common where there is some dust and/or marine particles \citep{2002JAtS...59..590D}.  The distribution from the current sky background model does not have a peak at larger radii.  This is because the current sky model used only a decomposition of the aerosol extinction curve of \citet{2011A&A...527A..91P}, which is not particularly sensitive to the presence of coarse grains, to find the distribution of aerosols.  Only with the observations at several different $\rho$, is it clear that coarse particles are needed.  This distinct lack of coarse mode particles in the current model could account for many of the differences seen between the current sky model and the optimal ones.  As can be seen in the \textit{bottom} panel, all of these values for the volume distribution are not well constrained and have typical uncertainties around 50\%.  Despite these large uncertainties, there seems to be clear tendencies between the optimal models and current model as well as between including dust or not.  The existence of a second peak from coarse modes is evident, however the location of the peak seen in the results depends on our choices of input parameters for the coarse mode distributions. 

\begin{table}
\caption{Most likely aerosol extinction curves \label{tab_ext}}
\centering
\begin{tabular}{cccccc}
\hline\hline
\noalign{\smallskip}
\,\,$\lambda$ \,\,\,\,\,\,\,\,\,\,\,\,\,\,\,\,\,\,\,\,\,\,\,\,\,\,\,\,\,\,\,\,\,\,\,\,\,\,\,\,\,\,\,\,\,&$\tau_\mathrm{aer}$ \,\,\,\,\,\,\,\,\,& & && \\
 \,\,$\mu$m\,\,\,\,\,\,\,\,\,\,\,\,\,\,\,\,\,\,\,\,\,\,\,\,\,\,\,\,\,\,\,\,\,\,\,\,\,\,\,\,\,\,\,\, &  $10^{-3}$\,\,\,\,\,\,\,\,\,\, & && &  \\
 \end{tabular}
 \begin{tabular}{ccccc}
  & Run A & Run A alt & Run B& Run C  \\
 \noalign{\smallskip}
\hline
\noalign{\smallskip}
     0.30 & 34(9)  & 27(7)  & 19(6)  & 39(14)\\
     0.35 & 33(8)  & 27(7)  & 18(6)  & 38(13)\\
     0.40 & 31(7)  & 25(6)   & 17(5) & 36(12)\\
     0.45 & 29(6)  & 23(5)  & 17(5)  & 33(10)\\
     0.50 & 27(6)  & 22(4)  & 16(4)  & 31(9)\\
     0.55 & 25(5)  & 20(4)  & 15(3)  & 28(8)\\
     0.60 & 23(4)  & 19(3)  & 14(3)  & 26(7)\\
     0.65 & 21(3)  & 17(3)  & 13(3)  & 24(6)\\
     0.70 & 20(3)  & 16(2)  & 13(3)  & 22(6)\\
     0.75 & 19(3)  & 15(2)  & 12(2)  & 21(6)\\
     0.80 & 18(3)  & 14(2)  & 12(2)  & 20(5)\\
     0.85 & 17(3)  & 14(2)  & 11(2)  & 19(5)\\
     0.90 & 16(3)  & 13(2)  & 11(2)  & 18(5)\\
     0.95 & 16(3)  & 13(2)  & 10(2)  & 17(5)\\
     1.00 & 15(3)  & 12(2)  & 10(2)  & 16(5)\\
     1.10 & 14(3)  & 11(2)  & 10(2)  & 16(5)\\
     1.20 & 14(3)  & 11(3)  & 9(3)  & 15(5)\\
     1.30 & 13(3)  & 11(3)  & 9(3)  & 14(5)\\
     1.40 & 13(3)  & 10(3)  & 9(3)  & 14(5)\\
     1.50 & 12(3)  & 10(3)  & 9(3)  & 13(5)\\
     1.60 & 12(3)  & 10(3)  & 9(3)  & 13(6)\\
     1.70 & 12(3)  & 10(3)  & 8(3)  & 13(6)\\
     1.80 & 12(3)  & 9(3)  & 8(3)  & 13(6)\\
     1.90 & 11(3)  & 9(3)  & 8(3)  & 12(6)\\
     2.00 & 11(3)  & 9(3)  & 8(3)  & 12(6)\\
\noalign{\smallskip}
\hline
\end{tabular}
\tablefoot{From the Bayesian analysis of the aerosol extinction curves at the zenith, showing the mean $<\tau_\mathrm{aer}>$ and the $1\,\sigma$ uncertainty in parenthesis.  These extinction values have also been plotted in Fig. \ref{best_ext}.}
\end{table}

\subsection{Caveat: 7$^\circ$ observations}

\begin{figure}[!ht]
   \centering
   \includegraphics[width=0.49\textwidth]{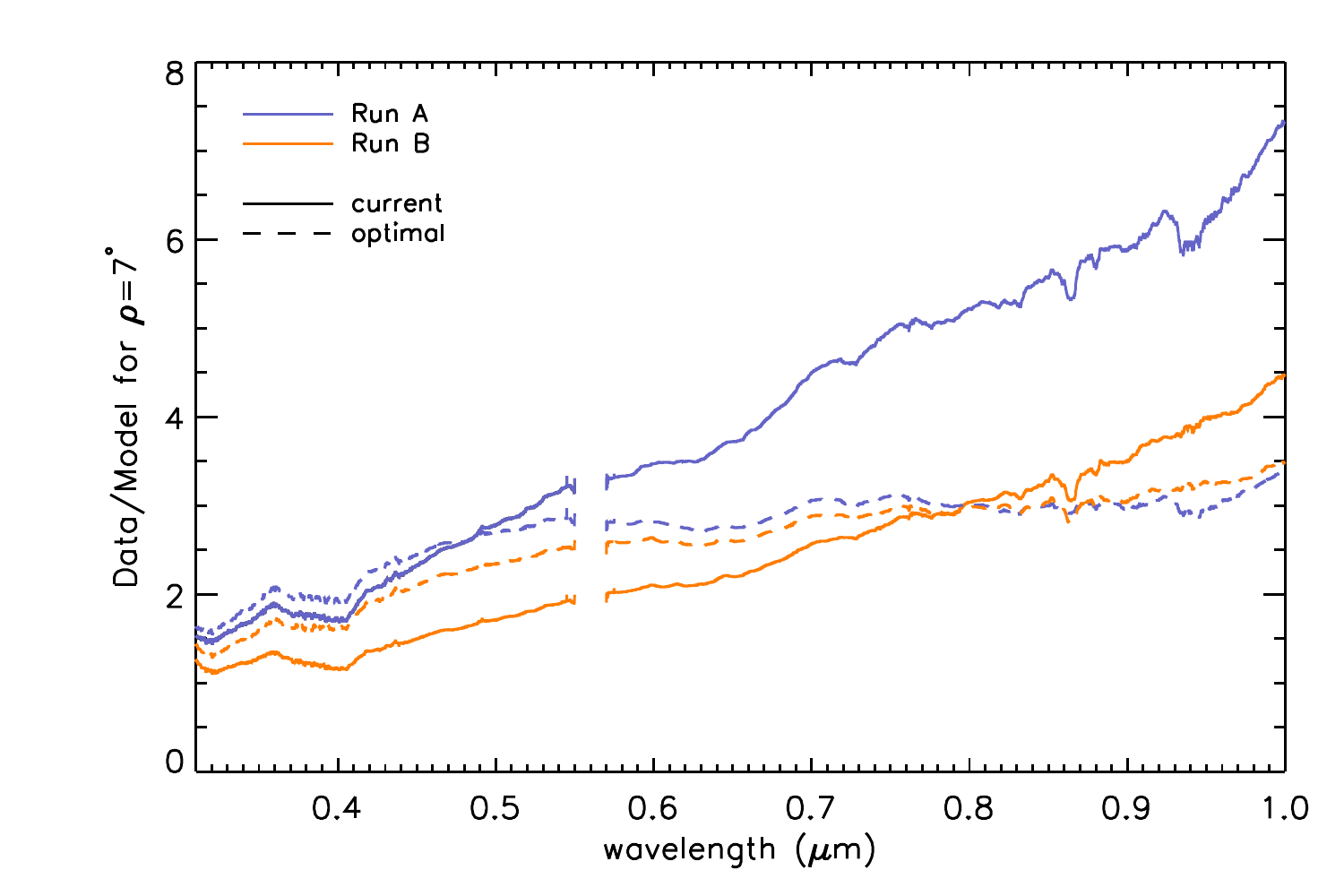}
   \caption{Shows the ratio of the observed data over the current model (solid) and optimal model (dashed), both with $F=1.0$, for Run A and B at $\rho=7^\circ$.  The ratio was smoothed to better show the trend with wavelength.  The gap is where the two arms overlap at the edges of the spectrographs and is clipped for clarity.}
\label{obs_7}
    \end{figure}

With the data at small $\rho$ there seems to be another source of scattered moonlight that is not taken into consideration by our current model.  In all three runs the $\rho=7^\circ$ observations had much more flux compared to the model.  In Fig. \ref{obs_7} the ratio of the observed flux over the modeled flux is plotted for Run A and B using both the current models and the most likely, optimal models.  Within the limits of our model, we tested many various aerosol distributions to reproduce this flux increase, including adding dust particles.  When only fitting the $\rho=7^\circ$, we could reproduce the observations, but with higher amounts of aerosols compared to the other observations (see Table \ref{tab_best_mod}).  Thus, we could not find an aerosol distribution that would consistently fit all 6 angular distances for a given night.  Therefore, this extra flux must be from some process that is beyond the scope of our current model.  This could be something specific to the X-Shooter instrument or a general effect that occurs when one observes close to the moon.

If this extra flux is specific to the X-Shooter instrument, then it could come from some internal scattering caused by a very bright object, the moon, close to the target.  There could also be scattering off the dome into the instrument.

There are many assumptions and limitations of our model that could lead to our inability to consistently fit all six observations on a given night.  We have assumed a constant refractive index $N'$ for all the aerosol particles.  $N'$ consists of a real and imaginary part and both can vary with wavelength and aerosol mode.  The imaginary part of $N'$ in remote areas like Cerro Paranal seems to be very small and should affect the absorption \citep{2002JAtS...59..590D}.  This effect could be from our choice of the single scattering albedo, which we also took as constant with respect with wavelength and aerosol type.  Both of these assumptions should mostly affect the wavelength dependence of our model and not the $\rho$ dependence.  We also assumed that all of the aerosol particles are spherically symmetric so we could make the Mie approximation.  This assumption appears to mostly affect the larger $\rho$ and not the smaller $\rho$ as seen by comparing scattering off of spheres and randomly orientated spheroids \citep[][and reference therein]{2002JAtS...59..590D}.  Additionally, we only tested a finite set of aerosol distributions which included varying parameters of remote background and dust aerosols.  Lastly, this could be a limitation in our scattering calculations.  The limitation could be regarding the number of grid points considered, or the multiple scattering corrections.  Also polarization is not considered in the scattering calculations.  It could also be from that the stratospheric aerosols are treated the same way as the tropospheric aerosols in the scattering calculations.

There are some suggestions that the extra flux seen at small $\rho$ may be a general phenomenon and that this may be evidence of a lunar aureole even in clear seeing conditions.  This type of an effect is also seen around bright stars \citep[e.g.][]{1971PASP...83..199K,1991PASP..103..645S} and the cause is currently unknown.  Solar and lunar aureole are known to occur when there are cirrus clouds or ice crystals present in the atmosphere and can extend several degrees from the sun or moon \citep[e.g.][]{2009JAtOT..26.2531D,2013JGRD..118.5679D,2017RSEnv.196..238R}.

Regardless of the origin of this extra flux, we strongly caution any optical observations with low signal-to-noise ratios at astronomical facilities that are taken close to the moon (within 10 or 15$^\circ$).

\section{Summary and Discussion}

We have a unique set of sky observations taken with X-Shooter to study the sky background, with a focus on the scattered moonlight and aerosol properties.  These observations were taken on three different nights, each with a different lunar phase.  Each night had a set of six observations taken at different angular distances $\rho$ from the moon.  This allowed us to test the current scattered moonlight model as a function of lunar phase and $\rho$, as well as the other components of the sky background model where the moonlight is not dominant.  Additionally, the current sky background model was limited in how well the aerosol properties could be determined.  The aerosol extinction curve was an extrapolation of one measured in the optical by \citet{2011A&A...527A..91P}, and the phase function was found by decomposing this extinction curve into various background aerosol distributions.  With the X-Shooter observations we could test a large suite of aerosol distributions from the UV to NIR.  For the first time we could study the effects of aerosols from the UV to NIR and at multiple $\rho$ to constrain the aerosol size distributions including the amount of coarse particles.

With the X-Shooter sky observations we were able to test the current scattered moonlight model.  When looking at parts of the spectrum that are dominated by Rayleigh scattering instead of aerosol scattering, we found that the current scattered moonlight model reflects the changes in lunar phase and $\rho$ well and the enhancement factor $F$ is consistent with one for this set of data.  Also, we noticed that there is still some contribution from scattered moonlight to continuum at redder wavelengths, e.g. in the J-band (see Appendix A for more details). 

For testing the full sky background model, we compared the model with observations at the redder wavelengths and larger $\rho$.  The continuum of the sky background model agrees well with the data, however the airglow continuum model tends to overestimate the amount of flux around $\lambda=1\,\mu$m.  This is one of the most uncertain components of the model due to the difficulty in measuring it as well as predicting its variability.

In order to determine better the aerosol extinction at Cerro Paranal from the UV to NIR, we created a set of sky background models with different aerosol distributions.  This way we could take advantage of using both the aerosol extinction curves and phase functions derived from each distribution to test the models against the observations.  By comparing the sections of the spectra containing only continuum, we found the best model for each observation through $\chi^2$ minimization.  The best-fit individual models were all in excellent agreement with the observations (except the 90 and 110$^\circ$ observations of Run A with cirrus clouds).  The observations with small $\rho$ tended towards higher amounts of aerosols.  Along with the remote background aerosols, we also included dust particles since Cerro Paranal is located in the desert.  Using a set of sky background models with different aerosol properties coupled with observations of the night sky is a new way to obtain the aerosol extinction curve and phase function at the time of observation.  There are several degeneracies amongst the different aerosol modes, however it seems to be more accurate than using the typical Langley method (which failed for our standard star observations at Cerro Paranal; see Section 3.2) and the uncertain decomposition of the resulting extinction curve into aerosol modes \citep[as][]{2013A&A...560A..91J}.

Using the likelihoods of each model from the $\chi^2$ minimization, we can perform a Bayesian analysis to find the most likely aerosol distribution for a given night.  Since there seemed to be extra flux at the $\rho=7^\circ$ observations that was difficult to fit consistently with the others, we ignore these for this analysis.  We found the most likely aerosol extinction curve, phase function, and aerosol volume density.  We then re-ran the sky background model with the most likely aerosol extinction curve and phase function, called the optimal model, for each run.  We compared these expressions and models with the current sky background model.  With this analysis using the set of X-Shooter observations, we are more sensitive to the different aerosol modes than the previous technique used for the current sky background model.  This previous technique of decomposing the optical extinction curve was mainly sensitive to the accumulation mode which dominates the shape of the extinction curve.  Now we are also sensitive to the amount of coarse modes.  We found that the optimal aerosol extinction curves vary some from night to night, but consistently have a different shape than the curve currently used in the sky background model.  Additionally, the most likely aerosol phase functions for all three runs are more peaked at small $\rho$ and are less smooth at large $\rho$ compared to the one used in the current sky background model.  Aerosols that are coarse mode particles appear to be present at Cerro Paranal and are an important factor for the aerosol phase function and volume densities.  

We can compare the changes in the optimal aerosol distributions for each run with  other data collected at Cerro Paranal.  There are some atmospheric data that are collected as part of the ESO ambient conditions and radiometer data \citep{2012SPIE.8446E..3NK}.  From the ESO ambient conditions website we can find information about the dust particle size for a given night and time of the ground layer at 20\,m.  The ground layer is below the X-Shooter instrument and can have large short time-scale variations.  It is unclear if these changes reflect changes in the troposphere.  The $0.5\,\mu$m\;dust at $20\,$m for the beginning and end of the observations for Run A was about 2.0$\times10^4$ and 6.9$\times10^4$\,m$^{-3}$, for Run B 1.4$\times10^4$ and 3.0$\times10^4$\,m$^{-3}$, and for Run C 2.2$\times10^5$ and 2.4$\times10^4$\,m$^{-3}$, respectively.  The estimated amount of dust during Run B was lower compared to the other two runs, especially the beginning of Run C and the end of Run A, which is consistent with what is seen in Figs. \ref{mod_pat} and \ref{best_ext}.  This also demonstrates how much the dust and aerosols can change in short time intervals.  With the radiometer data, we have estimates of the amount of integrated water vapor (IWV) for each run \citep{2012SPIE.8446E..3NK}.  At the beginning and end of Run A, the IWV was about 3.7 and 4.1\,kg\;m$^{-2}$, for Run B 1.6 and 1.7\,kg\;m$^{-2}$, and for Run C 2.4 and 2.6\,kg\;m$^{-2}$, respectively.  Again Run B was much drier compared to Run A and C, which is expected since Run B was done in winter.  These atmospheric data can provide some insight into the aerosol composition of a given night and time since the higher the humidity usually allows for larger aerosol particles.

We can also compare the optimal volume densities derived for Cerro Paranal with other similar sites observed by AERONET.  The most comparable site with extremely low aerosol optical depths is Mauna Loa Observatory, HI, USA \citep{2001JGR...10612067H}, however these observations were made during the daytime using the sun.  This site is located 3400\,m above sea level on the side of a shield volcano in the middle of the Pacific.  The combination of being at a high elevation, far away from any urban areas, and lack of vegetation allows for generally very low amount of aerosols.  The most frequent aerosol optical depth at 0.5\,$\mu$m measured is between 0.01 and 0.02 with a tail to higher optical depths, which is similar to our measurements at Cerro Paranal between 0.016 (Run B) and 0.031 (Run C).  Typical volume densities at Mauna Loa Observatory vary between having a single peak around a particle radius of 0.12\,$\mu$m and having another additional peak around 5\,$\mu$m.  The latter is most likely when some marine particles are blown upwards, which is similar to have a small amount of coarse particles at Cerro Paranal from the desert conditions.  There are two other AERONET stations that are in Chile and are the closest to Cerro Paranal, called Crucero and Cherenkov Telescope Array (CTA), which are about 275 and 12\,km away with altitudes of 1176 and 2154\,m, respectively.  Crucero is Northeast of Antofagasta, whereas Cerro Paranal is south of Antofagasta and Crucero is at much lower altitude compared to Cerro Paranal.  The conditions at Crucero and Cerro Paranal can be quite different.  Regardless, Crucero also can have low optical depth and a bi-modal peak in volume density but has only been operational since 2017.  The other site in the Atacama desert is also relatively new and is at the site of the CTA \citep{2017arXiv170807484J}.  They measured the aerosol optical depths using both the sun and direct measurements with the moon.  They generally also had low optical depths, with typical values between 0.01 and 0.02 at 0.5\,$\mu$m, which varied significantly during the day and month reaching values as high as 0.08 in their given example.  Our optimal values for the aerosol optical depths and volume densities for Cerro Paranal seem consistent with other sites with low amounts of aerosols as well as those sites that are located nearby. 

In general, accurately measuring the aerosol scattering for a given location and time is difficult, especially at Cerro Paranal where the aerosol optical depth is very low.  Additionally, the aerosol distributions and thus scattering properties can change on short and long time-scales.  To address the issue of very low optical depth, we introduce a new method for estimating the aerosol distributions by using a set of sky background models with different aerosol distributions coupled with sky observations with scattered moonlight.  This technique for finding the aerosol size distributions and hence the aerosol extinction for a given time and location could also be used for data reduction, however it requires the set of sky background models with a reasonable range of aerosol distributions as well as a set of sky observations with scattered moonlight and preferably at multiple $\rho$.  With only a single plain sky observation, this method would still work, but would have more degeneracies since the aerosol phase function would be less constrained.  This technique, even for a single observation, could be an improvement over typical methods that use spectrophotometric standard stars, which require accurate flux calibrations and well-known stars.

The most likely aerosol extinction curves, phase functions and volume densities given in this paper could improve the current sky background and correcting for the sky background at Cerro Paranal.  Indeed the aerosol properties amongst the three runs are different, however many characteristics are the same.  The general shapes of both the aerosol extinction curve and phase function are similar.  From the optimal volume densities it is clear that the aerosol distribution at Cerro Paranal is bi-modal with a peak at smaller and larger particles from the accumulation and coarse modes, respectively.  These larger particles affect the aerosol extinction curve at slightly longer wavelengths (around 1\,$\mu$m) and the shape of the phase function, causing the phase function to be more peaked at small $\rho$.  For a more accurate sky background estimate, these coarse modes should be included.

Overall the current sky background model is in agreement with the observations, with an enhancement factor $F=1.0$ and within the limitations of having an average aerosol scattering prescription.  From this work, including coarse particles could improve the accuracy of the current sky background model.  The amount of aerosols present at Cerro Paranal is very low.  The aerosol extinction values provided in Fig. \ref{best_ext} and Table \ref{tab_ext} can be used to improve the sky background model and data reduction.  These changes in aerosol properties amongst the runs would most likely have a small effect for most science targets, with the largest impact being for low signal-to-noise ratio observations between 0.5 and 1.0\,$\mu$m when the moon is above the horizon.  In the NIR, the aerosol extinction curve is practically negligible.  Since aerosols can vary on short timescales, in order to better measure the aerosol distributions for a given time, it is recommended that most major observatories install an instrument to simultaneously measure the aerosol scattering properties, like what is used in AERONET, which would additionally allow for more research on aerosol climatology. 

\section{Conclusion}

In this work we coupled a set of sky background models with different aerosol size distributions with X-Shooter observations of the plain sky to find the optimal aerosol size distribution at Cerro Paranal to both better characterize the night sky and to improve the sky background model.  Additionally, we could test the current sky background model, focusing mostly on the scattered moonlight component.  Here are the main conclusions and highlights of this study:

\begin{itemize}
\item The current scattered moonlight model reflects the changes in lunar phase and $\rho$ well;\\
\item The enhancement factor $F$ of 1.2 for the scattered moonlight in the original sky background model which was needed to match the FORS1 data, is not needed for the X-Shooter data and was changed in the ESO sky background model;\\
\item The first attempt to determine the nighttime aerosol size distributions by using a scattered moonlight model and a set of plain sky observations;\\
\item This new technique allowed for constraints on both the aerosol extinction curve and phase function;\\
\item The optimal aerosol extinction curves vary some from night to night, but consistently have a different shape than the curve currently used in the sky background model;\\
\item The most likely aerosol phase functions for all three runs are more peaked at small $\rho$ and are less smooth at large $\rho$ compared to the one used in the current sky background model;\\
\item The optimal volume densities for all three runs with and without dust show a bi-modal distribution of aerosols, a peak at smaller particles from the accumulation mode and a peak at larger particles from the coarse mode.  The current sky background model lacks the second peak of coarse particles;\\
\item The optimal volume densities are consistent with what is observed at other sites with very low aerosol optical depth.
\end{itemize}

\begin{acknowledgements}
This study used X-Shooter observations from the ESO proposal 491.L-0659.  It was carried out in the framework of the Austrian ESO In-kind project funded by BM:wf under contracts BMWF-10.490/0009-II/10/2009 and BMWF-10.490/0008-II/3/2011, and by the Austrian Science Fund (FWF) project P26130.  S.~Noll received funding from the project P26130 of the Austrian Science Fund (FWF) and is now financed by the project NO\,1328/1-1 of the German Research Foundation (DFG).  W.~Kausch is funded by the Hochschulraumstrukturmittel provided by the Austrian Federal Ministry of Education, Science and Research (BMBWF).  We thank B.~Holben and J.~Jury{\v s}ek for their efforts in establishing and maintaining Mauna Loa Observatory and Crucero, and CTA sites, respectively.  We also thank F.~Kerber (ESO) for providing us with the radiometer data for the three nights of observations.
\end{acknowledgements}

\bibliographystyle{aa}
\bibliography{moon_nir}

\clearpage
\begin{appendix}
\section{Scattered moonlight brightness in the J-band region}

Using the sky background model, we can estimate the amount of scattered moonlight in the NIR for different lunar phases and distances from the moon.  Usually when people observe in the NIR, they completely ignore any effect from scattered moonlight, since it is predominately in the optical.  However, as can be seen in Fig. \ref{eg_spec}, there is some scattered moonlight still present in the J-band wavelength region.  In this region the airglow emission is the dominant source of sky background noice.  However, if one was observing a faint source to study its continuum emission, scattered moonlight may need to be considered.    In order to quantify how much scattered moonlight one can expect at different distances from the moon and different lunar phases, we used the optimal models derived in Section 5.2 for each run.  We then considered only the scattered moonlight portion of the sky background flux.  We calculated the median flux from the scattered moonlight portion of the optimal models in a wavelength range with no strong airglow emission lines between 1.188 and 1.195\,$\mu$m, shown in the top half of Table \ref{tab_jband}.  Additionally, we provide the median fraction of scattered moonlight over the total sky background flux using the optimal models within the same narrow wavelength range, 1.188 to 1.195\,$\mu$m.  In the continuum part of the J-band, the fraction of scattered moonlight is still relatively high near full moon and when the moon is moderately close ($\lesssim45^\circ$) to the target.  When the moon is close to full and especially if the observations are near the moon, the scattered moonlight flux may not be negligible.  Also, note that the optimal models from Section 5.2 tend to underestimate the amount of flux at $\rho=7^\circ$, so the values provided are a conservative estimate.


\begin{table}
\caption{J-band continuum flux values and fractions of scattered moonlight \label{tab_jband}}
\centering
\begin{tabular}{ccccccc}
\hline\hline
\noalign{\smallskip}
Run & 7$^\circ$ & 13$^\circ$ & 20$^\circ$  & 45$^\circ$  & 90$^\circ$  & 110$^\circ$  \\
\noalign{\smallskip}
\hline
\noalign{\smallskip}
A   &   84  &   34  &   17  &   6.9  &  3.9  &  3.0\\
B   &   34  &   16  &   10  &   4.2  &  1.5  &  1.8\\
C   &   10  &   4.4  &   2.9  &   0.95  &  0.50  &  0.54\\
\noalign{\smallskip}
\hline
\noalign{\smallskip}
A   &   0.95  &   0.87  &   0.78  &   0.56  &  0.33  &  0.28 \\
B   &   0.90  &   0.78  &   0.69  &   0.47  &  0.25  &  0.23 \\
C   &   0.67  &   0.47  &   0.37  &   0.17  &  0.08  &  0.07 \\
\noalign{\smallskip}
\hline
\end{tabular}
\tablefoot{The top half is the median flux of scattered moonlight between 1.188 and 1.195\,$\mu$m in $10^2$\,photon\,s$^{-1}$m$^{-2}\mu$m$^{-1}$arcsec$^{-2}$.  The bottom half is the median fraction of scattered moonlight to the total sky background between 1.188 and 1.195\,$\mu$m.}
\end{table}

\end{appendix}

\end{document}